\documentclass[%
 aip,
 sd,%
 amsmath,amssymb,
preprint,%
 fleqn,
]{revtex4-1}

\usepackage{graphicx}
\usepackage{dcolumn}
\usepackage{bm}

\begin{document}


\title{Existence of dust ion acoustic solitary wave and double layer solution at $M = M_{c}$}
\author{Animesh Das}
\affiliation{ B. N. S. U. P. School, Howrah - 711 315, India.}%
\author{Anup Bandyopadhyay}
\email{abandyopadhyay1965@gmail.com}%
\affiliation{ Department of Mathematics, Jadavpur University, Kolkata
- 700032, India.}%
\author{K. P. Das}
\affiliation{ Department of Applied Mathematics, University of Calcutta, 92, Acharya Prafulla Chandra Road, Kolkata 700 009, India.}%
\date{\today}

\begin{abstract}
\noindent \noindent The Sagdeev potential technique has been used to investigate the existence and the polarity of dust ion acoustic solitary structures in an unmagnetized collisionless nonthermal dusty plasma consisting of negatively charged static dust grains, adiabatic warm ions and nonthermal electrons when the velocity of the wave frame is equal to the linearized velocity of the dust ion acoustic wave for long wave length plane wave perturbation, i.e., when the velocity of the solitary structure is equal to the acoustic speed. A compositional parameter space has been drawn which shows the nature of existence and the polarity of dust ion acoustic solitary structures at the acoustic speed. This compositional parameter space clearly indicates the regions for the existence of positive and negative potential dust ion acoustic solitary structures. Again, this compositional parameter space shows that the present system supports the negative potential double layer at the acoustic speed along a particular curve in the parametric plane. However, the negative potential double layer is unable to restrict the occurrence of all negative potential solitary waves. As a result, in a particular region of the parameter space, there exist negative potential solitary waves after the formation of negative potential double layer, i.e., negative potential supersolitons have been observed at the acoustic speed. But the amplitudes of these supersolitons are bounded. A finite jump between amplitudes of negative potential solitons separated by the negative potential double layer has been observed, and consequently, the present system supports the supersolitons at the acoustic speed in a neighbourhood of the curve along which negative potential double layer exist. The present system does not support any positive potential double layer for any physically admissible values of the parameters and consequently, the present system does not support any positive potential supersoliton at the acoustic speed. The effects of the parameters on the amplitude of the solitary structures at the acoustic speed have been discussed.  
\end{abstract}

\maketitle

\section{\label{sec:intro}Introduction}

In astrophysical environments highly (negative/positive) charged micronsize impurities or dust particulates are observed in electron-ion plasmas and such plasmas are found in supernovas, pulsar environments, cluster explosions, active galactic nuclei etc. The presence of dust grains having large masses introduces several new aspects in the properties of the nonlinear waves and coherent structures \cite{Rao90,Shukla92,Verheest92,Barkan96,Mamun96a,Mamun96b,Pieper96,Shukla01,Shukla09a,Shukla99,Shukla03}. Depending on different time scales, there can exist two or more acoustic waves in a typical dusty plasma. Dust Acoustic  (DA)  and  Dust Ion Acoustic  (DIA)  waves are  two  such  acoustic  waves. \citet{Shukla92} were the first to show that due to the quasi-neutrality condition $ n_{e0} + n_{d0}Z_{d}  = n_{i0} $ and the strong inequality $ n_{e0} \ll n_{i0} $ (where $ n_{e0}, n_{i0},$ and $ n_{d0} $ are, respectively, the number density of electrons, ions, and dust particles, and $ Z_{d} $ is the number of electrons residing  on  the  dust  grain  surface),  a  dusty  plasma (with negatively charged static dust grains) supports low-frequency Dust Ion Acoustic (DIA) waves with phase velocity much smaller (larger) than electron (ion) thermal velocity. In the case of  a  long  wavelength  limit,  the  dispersion  relation  of DIA wave is similar to that of IA wave for a plasma with $ n_{e0} = n_{i0} $ and $ T_{i} \ll T_{e} $, where $ T_{i}(T_{e}) $ is the average ion (electron) temperature. Due to the usual dusty plasma approximations ($ n_{e0} \ll n_{i0} $ and $ T_{i}  \simeq T_{e} $), a dusty plasma cannot support the usual IA waves, but a dusty plasma can support the DIA waves of \citet{Shukla92}. Thus, DIA waves are basically IA waves modified  by  the  presence  of  heavy  dust  particulates. The  theoretical  prediction  of  \citet{Shukla92} was supported by a number of laboratory experiments \cite{Barkan96,Merlino98,Nakamura01}. The nonlinear properties of DIA waves in different dusty plasma systems have been investigated by \citet{Bharuthram92} , \citet{Nakamura99} , \citet{Luo99} , \citet{Mamun02a} , \citet{Shukla03} ,  \citet{Verheest05} , \citet{Sayed08} , \citet{Alinejad10b}, \citet{Baluku10a} , \citet{das12}.

In most of the earlier works, the Maxwellian velocity distribution function for lighter species of particles has
been  used  to  study  DIA Solitary Waves (DIASWs) and  DIA  double  layers  (DIADLs).  However,  the  dusty  plasma  with  non-
thermally/suprathermally  distributed  particles  are  observed in a number of heliospheric environments (\citet{Asbridge68}; \citet{Feldman83}; \citet{Lundin89}; \citet{Verheest00}; \citet{Shukla02}; \citet{Futaana03}). In fact, the relaxation time for lighter species of particles is not so small to reach thermal equilibrium and hence different non-Maxwellian velocity distribution functions have been used (\citet{Djebli09}). Space plasma  observations  indicate  the  presence  of  ion  and electron populations, which are not in thermodynamic equilibrium. For example, energetic particles have been observed  in  and  around  the  Earth's  bowshock  and foreshock (\citet{Asbridge68}; \citet{Feldman83}) and  the  loss  of  energetic  particles  has  been  observed from the upper ionosphere of Mars (\citet{Lundin89}). Energetic  protons  have  been  observed  in  the  vicinity of the moon (\citet{Futaana03}). The model of the velocity distribution function of non-thermal electrons was considered for the first time by \citet{Cairns95a} to study the IA solitary structures in the presence of  the  population  of  fast  energetic  electrons  together with the population of Maxwellian distributed electrons. The other forms of distributions have also been used in  the  literature  to  characterize  non-thermal  features of particle distributions. The $\kappa$ - distribution is one such distribution. Therefore, it is of considerable importance to study nonlinear wave structures in a dusty plasma in which lighter species (electrons and / or ions and / or positrons) is non-thermally distributed. Such a study of dusty plasma consisting of Cairns non-thermal distributed ions has been made by several authors \cite{Mamun96b,Mendoza00,Maharaj04,Maharaj06,Verheest08,das09,das10}. \citet{berbri09} have investigated weakly nonlinear DIA shock waves in a  dusty  plasma  with  non-thermal  electrons.  \citet{Baluku10a}  have  investigated  DIASWs  in an unmagnetized dusty plasma consisting of cold dust particles  and  kappa  distributed  electrons  using  both small and arbitrary amplitude techniques. The nonlinear theory of DIA waves in an unmagnetized collisionless nonthermal dusty plasma consisting of negatively charged static dust grains, adiabatic warm ions and nonthermal electrons has been investigated by \citet{das12} when the velocity of the wave frame is strictly greater than the linearized velocity of the dust ion acoustic wave for long wave length plane wave perturbation, i.e., when the velocity of the solitary structure is strictly greater than the acoustic speed.

In most of the earlier works \cite{Shukla78, Yu80, Bharuthram86, Baboolal88, Baboolal89, Baboolal90, Baboolal91, Cairns95a, Cairns95b, Popel96, Mamun96a, Mamun96b, Xie98, Mendoza00, Gill03, Maharaj04, Maharaj06, Hellberg06, Verheest08, Baluku08, Tanjia08, Verheest09, Verheest10a, Djebli09, das09, das10} solitary waves and/or double layers have been investigated for $U>C_{D}$, where $U$ is the velocity of the wave frame and $C_{D}$ is linearized velocity of the dust ion acoustic wave for long wave length plane wave perturbation. However, some investigations \cite{Verheest10b, Baluku10a, Baluku10b} have shown that finite amplitude solitary wave can exist at $U=C_{D}$ in the parameter regime where solitons of both polarities exist. The numerical observations \cite{Verheest10b, Baluku10a, Baluku10b} of the solitary wave solution of the well known energy integral at $U=C_{D}$, influenced \citet{das12} to set up a general analytical theory for the existence and the polarity of solitary wave and double layer solution of the energy integral at $U=C_{D}$. The existence of the solitary waves and/or double layers at the acoustic speed have been analytically investigated by \citet{das12}. \citet{das12} have used the \citet{Sagdeev66} potential techniques to investigate the existence and the polarity of the solitary waves and/or double layers at the acoustic speed. In the present paper, we develop a computational scheme to investigate the existence and the polarity of Dust Ion Acoustic (DIA) solitary structures at the acoustic speed, in an unmagnetized nonthermal plasma consisting of negatively charged static dust grains, adiabatic warm ions and nonthermal electrons.

Supersolitons is a new class of solitons having some special characteristics along with the properties of traditional solitons. One can define supersoliton in the following way: we know that a sequence of positive (negative) potential solitary waves having monotonically increasing amplitude converges to a positive (negative) potential double layer if it exists, i.e., existence of a positive (negative) potential double layer implies that the existence of at least one sequence of positive (negative) potential solitary waves having monotonically increasing amplitude converging to that double layer solution. If there exists a parameter regime for which the  positive (negative) potential double layer is unable to restrict the occurrence of all positive (negative) potential solitary waves then there exist positive (negative) potential solitary waves after the formation of the positive (negative) potential double layer. The positive (negative) potential solitary wave after the formation of the positive (negative) potential double layer is known as the supersoliton.   \citet{Dubinov12} used the term `supersolitons' for the first time. Some of the special properties of supersolitons have been reported in the earlier papers (\cite{White72,Verheest09,Baluku10b,Verheest11,das12,Hellberg13,Verheest13a,Verheest13b,Verheest13c}). For example, \citet{das12} clearly state the following property for the existence of the supersolitons: supersolitons can exist if there exist two types solitary waves of same polarity separated by a double layer of same polarity. However, recently, \citet{Verheest14} critically analyzed electrostatic supersolitons in dusty plasmas mentioning various characteristic of supersolitons. However, in the most of the papers, supersolitons have been investigated for $U>C_{D}$. In the present paper, we develop a computational scheme to investigate the existence and the polarity of Dust Ion Acoustic (DIA) supersolitons at the acoustic speed, i.e., when $U=C_{D}$ in an unmagnetized nonthermal plasma consisting of negatively charged static dust grains, adiabatic positive ions and nonthermal electrons.

In the present investigation we have considered the problem of existence as well as the polarity of Dust Ion Acoustic Solitary Waves (DIASWs) and Dust Ion Acoustic Double Layers (DIADLs) in a nonthermal dusty plasma consisting of negatively charged dust grains, adiabatic warm ions and nonthermal electrons at the smallest possible value of the Mach number $M$, where the Mach number ($M$) or equivalently, the dimensionless velocity ($M$) of the wave frame is defined by $M=\displaystyle \frac{U}{(L/T)}$, $L$ is a characteristic length and $T$ is a characteristic time. Therefore, if $M_{c}= \displaystyle \frac{C_{D}}{(L/T)}$, $M_{c}$ is the smallest possible value of the Mach number $M$ and $U=C_{D} \Leftrightarrow M=M_{c}$, and consequently, our aim is to investigate the existence and the polarity of DIA solitary structures when $M=M_{c}$. The Sagdeev potential approach [\citet{Sagdeev66}] has been considered to investigate the existence of solitary wave and double layer at $M=M_{c}$. Three basic parameters of the present dusty plasma system are $\mu$, $\alpha$ and $ \beta_{1}$, which are respectively the ratio of unperturbed number density of nonthermal electrons to that of ions, the ratio of average temperature of ions to that of nonthermal electrons, a parameter associated with the nonthermal distribution of electrons. Nonthermal distribution of electrons becomes isothermal one if $\beta_{1}=0$ and consequently, for isothermal electron species, the present dusty plasma contains only two basic parameters $\mu$ and $\alpha$. Depending on the nature of existence DIA solitary structures, we have three cut off values $\mu_{p}$, $\mu_{c}$, and $\mu_{r}$ of $\mu$ and three cut off values $\beta_{1c}$, $\beta_{c}$ and $\beta_{1a}$ of $\beta_{1}$ such that the entire $\mu \beta_{1}$ parametric plane can be delimited into different regions of existence of DIA solitary structures at $M=M_{c}$. For any fixed value of $\alpha$, we have the following observations with respect to the cut off values $\mu$ and $\beta_{1}$:

\begin{itemize}
	\item $0<\mu<\mu_{p}$ : The system does not support any solitary structure at $M=M_{c}$ for any admissible values of $\beta_{1}$, i.e., for any $\beta_{1}$ lying within $0 \leq \beta_{1} \leq \beta_{1T}$.
	\item $\mu_{p} \leq \mu < \mu_{c}$ :	
\begin{enumerate}
	\item The system supports positive potential solitary structure at $M=M_{c}$ for any $\beta_{1}$ lying within $0 \leq \beta_{1} \leq \beta_{1a}$.
	\item The system does not support any solitary structure at $M=M_{c}$ for any $\beta_{1} > \beta_{1a}$.
\end{enumerate}
  \item $\mu_{c} < \mu \leq \mu_{r}$ :
\begin{enumerate}
	\item The system supports negative potential solitary structure at $M=M_{c}$ for any $\beta_{1}$ lying within $0 \leq \beta_{1} < \beta_{c}$.
	\item The system supports positive potential solitary structure at $M=M_{c}$ for any $\beta_{1}$ lying within $\beta_{c} < \beta_{1} \leq \beta_{1a}$.
	\item The system does not support any solitary structure at $M=M_{c}$ for any $\beta_{1} > \beta_{1a}$.
\end{enumerate}
	 \item $\mu_{r} < \mu \leq \mu_{T}$ :
\begin{enumerate}
	\item The system supports negative potential solitary structure at $M=M_{c}$ for any $\beta_{1}$ lying within $0 \leq \beta_{1} < \beta_{1c}$.
	\item The system supports negative potential double layer at $M=M_{c}$ for $\beta_{1} = \beta_{1c}$.
	\item The system supports negative potential solitary structure at $M=M_{c}$ for any $\beta_{1}$ lying within $\beta_{1c} < \beta_{1} < \beta_{c}$.
	\item The system supports positive potential solitary structure at $M=M_{c}$ for any $\beta_{1}$ lying within $\beta_{c} < \beta_{1} \leq \beta_{1a}$.
	\item The system does not support any solitary structure at $M=M_{c}$ for any $\beta_{1} > \beta_{1a}$.
\end{enumerate}
\end{itemize}
For any fixed value of $\alpha$ and for any $\mu$ lying within the interval $\mu_{r} < \mu \leq \mu_{T}$, we see from item 1., item 2. and item 3. that negative potential solitary waves at the acoustic speed within the interval $0 < \beta_{1} \leq \beta_{c}$ is separated by the negative potential double layer at $\beta_{1} = \beta_{1c}$. On the other hand, at the acoustic speed, there exists negative potential solitary waves after the formation of negative potential double layer at $\beta_{1} = \beta_{1c}$, and consequently, for any given value of $\alpha$ and for any $\mu$ lying within the interval $\mu_{r} < \mu \leq \mu_{T}$, the negative potential solitary structure at $M=M_{c}$ for any $\beta_{1}$ lying within $\beta_{1c} < \beta_{1} \leq \beta_{c}$ is actually a negative potential supersoliton at the acoustic speed.

Again, from the above observations, we see that there does not exist any solitary wave or any double layer at the acoustic speed when $\beta_{1} = \beta_{c}$ and / or $\mu = \mu_{c}$. We shall see later that when $\beta_{1}$ takes the value $\beta_{c}$ then $\mu$ assumes the value $\mu_{c}$ and conversely. Again, we have seen that positive (negative) solitons collapse at $\beta_{1} = \beta_{c}$ ($\Leftrightarrow \mu = \mu_{c}$). In fact, we have found an interesting phenomena around the point $\beta_{1} = \beta_{c}$. The amplitude of NPSW at the acoustic speed decreases with increasing $\beta_{1}$ and ultimately, demolished at $\beta_{1} = \beta_{c}$. On the other hand, PPSW at the acoustic speed starts to exist for $\beta_{1} > \beta_{c}$ and the amplitude of PPSW increases with increasing $\beta_{1}$, having maximum amplitude at $\beta_{1} = \beta_{1a}$. Therefore, the point $\beta_{1} = \beta_{c}$ acts as sink for NPSWs at the acoustic speed whereas the same point acts as a source for PPSWs at the acoustic speed.

The present paper is organized as follows: In \S \ref{sec:non eqn}, the basic equations are given. The energy integral and the Sagdeev potential has been constructed in \S \ref{sec:energy_int}. In \S \ref{sec:theory}, an analytical theory has been presented to determine the polarity of the solitary waves at the lowest value of the mach number. In the same section, the existence of the solitary structures at the critical value of the Mach number have been confirmed through the general analytical theory of \citet{das12mc} along with the use of the compositional parameter space showing the nature of existence of solitary structrues in a right neighbourhood of the curve $M=M_{c}$. Finally, a brief summary along with the discussions have been given in \S \ref{sec:conclusion}.

\section{\label{sec:non eqn}Basic equations}

The following are the governing equations describing the non-linear behaviour of dust ion acoustic waves  propagating  along  x-axis  in  collisionless  unmagnetized  dusty  plasma  consisting of adiabatic warm ions,  negatively  charged  immobile  dust  grains, and non-thermally distributed electrons:
\begin{eqnarray}\label{continuity}
\frac{\partial n_{i}}{\partial t}+\frac{\partial}{\partial x}(n_{i}u_{i})=0,
\end{eqnarray}
\begin{eqnarray}\label{momentum}
n_{i}m_{i}\bigg(\frac{\partial u_{i}}{\partial t}+u_{i}\frac{\partial u_{i}}{\partial x}\bigg)+\frac{\partial p_{i}}{\partial x}+n_{i}q_{i}\frac{\partial \phi}{\partial x}=0,
\end{eqnarray}
\begin{eqnarray}\label{pressure}
\frac{\partial p_{i}}{\partial t}+u_{i}\frac{\partial p_{i}}{\partial x}+\gamma p_{i} \frac{\partial u_{i}}{\partial x}=0,
\end{eqnarray}
\begin{eqnarray}\label{poisson}
\frac{\partial^{2} \phi}{\partial x^{2}}=-4\pi e(n_{i}-n_{e}-Z_{d}n_{d}).
\end{eqnarray}

\noindent Here $ n_{i}$, $n_{e}$, $n_{d}$, $u_{i}$, $p_{i}$, $\phi$, $x $ and $ t $ are, respectively, ion number density, electron number density, dust particle number density, ion fluid velocity, ion fluid pressure, electrostatic potential, spatial variable and time, $ \gamma(=3) $ is the adiabatic index, $ m_{i} $ is the mass of ion fluid, $ Z_{d} $ is the number of negative unit charges residing on the dust grain surface and $ e $ is the charge of an electron.

\noindent The above equations are supplemented by nonthermally distributed electrons as prescribed by \citet{Cairns95a} for the electron species. Actually, in a number of heliospheric environments, dusty plasma contains nonthermally distributed ions or electrons. Therefore, it is of considerable importance to study DIASWs and DIADLs in dusty plasmas in which electrons are nonthermally distributed. Nonthermal distribution of any lighter species of particles (as prescribed by \citet{Cairns95a} for the electron species) can be regarded as population of Boltzmann distributed particles together with a population of energetic particles. This can also be regarded as a modified Boltzmann distribution, which has the property that the number of particles in phase space in the neighbourhood of the point $v=0$ is much smaller than the number of particles in phase space in the neighbourhood of the point $v=0$ for the case of Boltzmann distribution, where $v$ is the velocity of the particle in phase space. This type of velocity distribution is often termed as Cairns distribution and was considered by many authors in various studies of different collective processes in plasmas and dusty plasmas \cite{Cairns95b, Mamun96c, Mamun96a, Bandyopadhyay99, Bandyopadhyay00a, Bandyopadhyay00b, Mendoza00, Bandyopadhyay01a, Bandyopadhyay01b, Bandyopadhyay02a, Bandyopadhyay02b, Bandyopadhyay02c, Gill03, Maharaj04, Maharaj06, Djebli09, das09, das10, Verheest08, Verheest10a, Verheest10b, das12}. Following \citet{Cairns95a}, the number density of non-thermal electrons can be written as
\begin{eqnarray}\label{ne}
\frac{n_{e}}{n_{e0}}=\bigg[1-\beta_{1}\frac{\phi}{\Phi}+\beta_{1}\bigg(\frac{\phi}{\Phi}\bigg)^{2}\bigg]exp\bigg[\frac{\phi}{\Phi}\bigg]
\end{eqnarray}
where
\begin{eqnarray}\label{beta_e}
\beta_{1}=\frac{4\alpha_{1}}{1+3\alpha_{1}}~,~\mbox{with}~\alpha_{1} \geq 0,
\end{eqnarray}
\begin{eqnarray}\label{Phi}
\Phi=\frac{K_{B}T_{e}}{e}.
\end{eqnarray}
Here $n_{e0}$ is the unperturbed electron number density, $K_{B}$ is the Boltzmann constant, and $T_{e}$ is the average temperature of electrons.  Here $ \beta_{1}$ ($\alpha_{1}$) is the parameter associated with non-thermal distribution of electrons and this parameter determines  the  proportion  of  fast  energetic  electrons. From the equation (\ref{beta_e}) and the inequality $\alpha_{1}\geq 0 $, it can be easily checked that the non-thermal parameter $ \beta_{1} $  is restricted by  the  inequality:  $ 0 \leq \beta_{1} < 4/3 $.  However, we cannot take the whole region of  $\beta_{1}$   (i.e., $ 0 \leq \beta_{1} < 4/3 $).  Plotting the non-thermal velocity distribution of electrons against its velocity ($ v $) in phase space, it can be easily shown that the number of electrons in phase space in the neighborhood of the point $v = 0$ decreases with  increasing   $ \beta_{1} $   and  the  number  of  electrons  in phase  space  in  the  neighborhood  of  the  point $ v  = 0 $ is almost zero when  $ \beta_{1} \rightarrow 4/3 $.  Therefore, for increasing $ \beta_{1} $, the distribution function develops wings, which become stronger for large value of $ \beta_{1} $ , and at the same time the center density  in  phase  space  drops,  the  latter  as  a  result of  the  normalization  of  the  area  under  the  integral. Consequently, we should not take values of $ \beta_{1} > 4/7 $ as that stage might stretch the credibility of the Cairns model too far [\citet{Verheest08}].  So, here we consider the effective range of $\beta_{1}$  as follows:
$ 0 \leq \beta_{1} \leq \beta_{T} $, where $\beta_{T}  = 4/7 \approx 0.571429 $.

\noindent Now introducing a new parameter
\begin{equation}\label{mu}
\mu = \frac{n_{e0}}{n_{i0}},
\end{equation}
the charge neutrality condition,
\begin{equation}\label{charge neutrality without mu}
Z_{d} n_{d0}+n_{e0}=n_{i0},
\end{equation}
can be written as
\begin{equation}\label{charge neutrality with mu}
\frac{Z_{d} n_{d0}}{n_{i0}} = 1 -\mu,
\end{equation}
where $n_{e0}$, $n_{i0}$ and $n_{d0}$ are, respectively, the unperturbed number densities of electron, ion and dust particulate.
\section{\label{sec:energy_int} Energy Integral}
Now introducing another parameter
\begin{eqnarray}\label{alpha}
\alpha=\frac{T_{i}}{T_{e}},
\end{eqnarray}
the linear dispersion relation of the DIA wave for the present dusty plasma system can be written as
\begin{eqnarray}\label{dispersion_relation}
\frac{\omega}{k}=C_{D}\sqrt{\frac{1+\frac{\gamma \alpha}{M_{s}^{2}}k^{2}\lambda_{D}^{2}}{1+k^{2}\lambda_{D}^{2}}},C_{D}=C_{s}M_{s},
\end{eqnarray}
where $\omega$ and $k$ are respectively the wave frequency and wave number of the plane wave perturbation, and
\begin{eqnarray}\label{cD}
C_{s}=\sqrt{\frac{K_{B}T_{e}}{m_{i}}}, M_{s}=\sqrt{\gamma\alpha+\frac{1}{\mu(1-\beta_{1})}},
\end{eqnarray}
\begin{eqnarray}\label{lambda_D_square}
\frac{1}{\lambda_{D}^{2}}=\frac{(1-\beta_{1})\mu}{\lambda_{Dem}^{2}}~,~
\lambda_{Dem}^{2}=\frac{K_{B}T_{e}}{4\pi e^{2}n_{i0}}.
\end{eqnarray}

Now for long-wave length plane wave perturbation, i.e., for $ k \rightarrow 0 $, from linear dispersion relation (\ref{dispersion_relation}), we have,
\begin{eqnarray}
\lim_{k \to 0}\frac{\omega}{k} =C_{D} \mbox{   and   } \lim_{k \to 0}\frac{d\omega}{dk} = C_{D}
\end{eqnarray}
and consequently the dispersion relation (\ref{dispersion_relation}) shows that the linearized velocity of the DIA wave in the present plasma system is $ C_{D} $ with $ \lambda_{D} $ as the Debye length.

To  study  the  arbitrary  amplitude  time  independent DIA solitary waves and double layers, we make all the dependent variables depend only on a single variable $ \xi=x-Ut $ where $ U $ is independent of $ x $ and $ t $. Thus, in the wave frame moving with a constant velocity $U$ the equations (\ref{continuity})-(\ref{poisson}) can be put in the following form
\begin{eqnarray}\label{modified_continuity}
\frac{d}{d\xi}\bigg\{(U-u_{i})\frac{n_{i}}{n_{i0}}\bigg\}=0,
\end{eqnarray}
\begin{eqnarray}\label{modified_momentum}
\frac{d}{d\xi}\big(-Uu_{i}+\frac{u_{i}^{2}}{2}+C_{s}^{2}\frac{\phi}{\Phi}\big)
 +\sigma_{ie}C_{s}^{2}\frac{n_{i0}}{n_{i}}\frac{d}{d\xi}\big(\frac{p_{i}}{P}\big)=0,
\end{eqnarray}
\begin{eqnarray}\label{modified_pressure}
\frac{d}{d\xi}\big\{(U-u_{i})\big(\frac{p_{i}}{P}\big)^{1/\gamma}\big\}=0,
\end{eqnarray}
\begin{eqnarray}\label{modified_poisson}
\frac{d^{2}}{d\xi^{2}}\bigg(\frac{\phi}{\Phi}\bigg) = \frac{1}{\lambda_{Dem}^{2}}\Bigg[1-\mu+ \mu\frac{n_{e}}{n_{e0}}-\frac{n_{i}}{n_{i0}}\Bigg].
\end{eqnarray}
Here $ P=n_{i0}K_{B}T_{i} $, $ n_{i0} $ is the unperturbed ion number density and $ T_{i} $ is the average temperature of ions.

Using the boundary conditions,
\begin{eqnarray}\label{boundary_condition}
\big(\frac{n_{i}}{n_{i0}},\frac{p_{i}}{P},u_{i},\phi,\frac{d\phi}{d\xi}\big)\rightarrow \big(1,1,0,0,0\big)\mbox{    as    }  |\xi|\rightarrow \infty
\end{eqnarray}
 and solving (\ref{modified_continuity}), (\ref{modified_momentum}), and (\ref{modified_pressure}), we get a quadratic equation  for  $ n_{i}^{2} $, and  the  solution of the final equation of $ n_{i} $ can be put in the following form:
\begin{eqnarray}\label{ni_square}
\frac{n_{i}}{n_{i0}}=N_{i}=\frac{(U/C_{s})\sqrt{2}}{\sqrt{\frac{\Phi_{U}}{\Phi}-\frac{\phi}{\Phi}}+\sqrt{\frac{\Psi_{U}}{\Phi}-\frac{\phi}{\Phi}}},
\end{eqnarray}
where
\begin{eqnarray}\label{phi_u}
\frac{\Phi_{U}}{\Phi} = \frac{1}{2}\big(\frac{U}{C_{s}}+\sqrt{3\sigma_{ie}}\big)^{2},
\frac{\Psi_{U}}{\Phi} = \frac{1}{2}\big(\frac{U}{C_{s}}-\sqrt{3\sigma_{ie}}\big)^{2}
\end{eqnarray}
Now  integrating  (\ref{modified_poisson})  with  respect  to  $ \phi $  and  using the  boundary  conditions  (\ref{boundary_condition}),  we  get  the  following equation  known  as  energy  integral  with  $ W(\phi) $  as  the Sagdeev potential or pseudo-potential:
\begin{eqnarray}\label{energy_integral}
\frac{1}{2}\bigg(\frac{d\phi}{d\xi}\bigg)^{2}+W(\phi)=0
\end{eqnarray}
where
\begin{eqnarray}\label{V_phi}
W(\phi) = \frac{\Phi^{2}}{\lambda_{Dem}^{2}}\Bigg[~W_{i} - \mu W_{e}-(1-\mu)W_{d}\Bigg],
\end{eqnarray}
\begin{eqnarray}\label{V_i}
W_{i} = \frac{U^{2}}{C_{s}^{2}}+\sigma_{ie}- N_{i}\big(\frac{U^{2}}{C_{s}^{2}}+3\sigma_{ie}
-2\frac{\phi}{\Phi}-2\sigma_{ie}N_{i}^{2}\big),
\end{eqnarray}
\begin{eqnarray}\label{V_e}
W_{e}=\bigg(1+3\beta_{1}-3\beta_{1}\frac{\phi}{\Phi}
+\beta_{1}\frac{\phi^{2}}{\Phi^{2}}\bigg)exp\bigg(\frac{\phi}{\Phi}\bigg)-(1+3\beta_{1}),
\end{eqnarray}
\begin{eqnarray}\label{V_d}
W_{d} = \frac{\phi}{\Phi}.
\end{eqnarray}

\noindent Using the mechanical analogy, \citet{Sagdeev66} established that for the existence of a Positive Potential Solitary Wave (Negative Potential Solitary Wave) solution of the energy integral (\ref{energy_integral}), the following three conditions must be simultaneously satisfied.
\begin{description}
  \item[Sa :] $\phi=0$ is the position of unstable equilibrium of a particle of unit mass associated with the energy integral (\ref{energy_integral}) under the action of the force field $-W'(\phi)$, i.e., $W(0)=W'(0)=0$ and $W''(0)<0$.
  \item[Sb :] $W(\phi_{m}) = 0$, $W'(\phi_{m}) > 0$ for some $\phi_{m} > 0$ ($W'(\phi_{m}) < 0$ for some $\phi_{m} < 0$). This condition is responsible for the oscillation of a particle of unit mass associated with the energy integral (\ref{energy_integral}) under the action of the force field $-W'(\phi)$ within the interval $\min\{0,\phi_{m}\}<\phi<\max\{0,\phi_{m}\}$.
  \item[Sc :] $W(\phi) < 0$ for all $0 <\phi < \phi_{m}$ ($W(\phi) < 0$ for all $\phi_{m} < \phi < 0$). This condition is necessary to define the energy integral (\ref{energy_integral}) as the equation of energy of a particle of unit mass moving along a straight line whose position is $\phi$ at time $\xi$ with velocity $\frac{d \phi}{d \xi}$ under the action of the force field $-W'(\phi)$ within the interval $\min\{0,\phi_{m}\}<\phi<\max\{0,\phi_{m}\}$.
\end{description}

On the other hand, if $W'(\phi_{m}) = 0$ along with the condition $W(\phi_{m}) = 0$ then the velocity and the force acting on the particle at $\phi =\phi_{m}$ are simultaneously equal to zero and consequently the particle will not be reflected back again at $\phi=0$. In this case instead of soliton solution Energy Integral (\ref{energy_integral}) gives shock-like solution which is known as double layer solution. If $\phi_{m} > 0$ the double layer is known as positive potential double layer (PPDL) whereas if $\phi_{m} < 0$ the double layer is known as negative potential double layer (NPDL). Therefore, for the existence of a Positive Potential Double Layer (Negative Potential Double Layer) solution of the energy integral (\ref{energy_integral}), the following three conditions must be simultaneously satisfied.
\begin{description}
  \item[Da :] $\phi=0$ is the position of unstable equilibrium of a particle of unit mass associated with the energy integral (\ref{energy_integral}) under the action of the force field $-W'(\phi)$, i.e., $W(0)=W'(0)=0$ and $W''(0)<0$.
  \item[Db :] $W(\phi_{m}) = 0$, $W'(\phi_{m}) = 0$, $W''(\phi_{m}) < 0$ for some $\phi_{m} > 0$ ($\phi_{m} < 0)$. This condition actually states that the particle cannot be reflected back again at $\phi = 0$.
  \item[Dc :] $W(\phi) < 0$ for all $0 <\phi < \phi_{m}$ ($W(\phi) < 0$ for all $\phi_{m} < \phi < 0$). This condition is necessary to define the energy integral (\ref{energy_integral}) as the equation of energy of a particle of unit mass moving along a straight line whose position is $\phi$ at time $\xi$ with velocity $\frac{d \phi}{d \xi}$ under the action of the force field $-W'(\phi)$ within the interval $\min\{0,\phi_{m}\}<\phi<\max\{0,\phi_{m}\}$.
\end{description}

Therefore, the  necessary  conditions  for  the  existence  of  solitary structures  of  the  energy  integral  (\ref{energy_integral})  are  $W (0) = 0$, $W '(0)  =  0$,  and $ W ''(0)  <  0$.  It  can  be  easily  checked that $W (0) = 0$, $W '(0)  =  0$, and the condition $ W ''(0)  <  0$ gives $ U  > C_{D} $. Our aim is to investigate the solitary structures when $ U  = C_{D} $. To investigate the solitary structures at the acoustic speed, i.e., at $ U  = C_{D} $,   we  have  introduced the dimensionless velocity $M$ of the wave frame as: $ M = U/C_{s}$. Now $ U  > C_{D} \Leftrightarrow U > C_{s}M_{s} \Leftrightarrow \frac{U}{C_{s}} > M_{s} \Leftrightarrow M > M_{s}$, and consequently, in general, the solitary structures start to exist for $M > M_{c}=M_{s}$ but our Our aim is to investigate the solitary structures when $ U  = C_{D} \Leftrightarrow U = C_{s}M_{s} \Leftrightarrow \frac{U}{C_{s}} = M_{s} \Leftrightarrow M = M_{s}$.  So  we  have  introduced  the following dimensionless quantities: $ \bar{x}=x/\lambda_{Dem} $, $ \overline{\xi}=\xi/\lambda_{Dem} $, $ \overline{t}=t/(\lambda_{Dem}/C_{s}) $, $ \overline{u_{i}}=u_{i}(\lambda_{Dem}/C_{s})/\lambda_{Dem} = u_{i}/C_{s} $, $ \overline{\phi}=\phi/\Phi $, $ \overline{p_{i}}=p_{i}/P $, $ \overline{n_{s}}=n_{s}/n_{i0} $. Now we note the following fact: $\xi=x-Ut=\lambda_{Dem}\overline{x}-U\frac{\lambda_{Dem}}{C_{s}}\overline{t}=\lambda_{Dem}[\overline{x}-\frac{U}{C_{s}}\overline{t}]
=\lambda_{Dem}[\overline{x}-M\overline{t}] \Leftrightarrow \frac{\xi}{\lambda_{Dem}}=\overline{x}-M\overline{t} \Leftrightarrow \overline{\xi}=\overline{x}-M\overline{t}$, i.e., here spatial coordinate is normalized by $\lambda_{Dem}$ and the time is normalized by $\frac{\lambda_{Dem}}{C_{s}}$. Then with respect to these dimensionless quantities, the energy integral can be simplified as follows, where we drop 'overline' on both independent and dependent variables:

\begin{eqnarray}\label{energy int}
\frac{1}{2} \bigg(\frac{d\phi}{d\xi}\bigg)^{2}+V(\phi) = 0,
\end{eqnarray}
where
\begin{eqnarray}\label{v(phi)}
    V(\phi) \equiv V(M,\phi,\beta_{1},\mu,\alpha) =  V_{i}-\mu V_{e}  - (1-\mu) \phi,
\end{eqnarray}
\begin{eqnarray}\label{V_i_mod}
    V_{i} = M^{2} + \alpha -N_{i} (M^{2}+3\alpha-2 \phi-2 \alpha N_{i}^{2}),
\end{eqnarray}
\begin{eqnarray}\label{V_e non}
    V_{e} = (1+3\beta_{1}-3 \beta_{1}\phi+\beta_{1}\phi^{2})e^{\phi}-(1+3\beta_{1}),
\end{eqnarray}
\begin{eqnarray}\label{n_i with Phi_M Psi_M 2}
    N_{i} = \frac{\sqrt{2}M}{\sqrt{\Psi(M)-\phi}+\sqrt{\Phi(M)-\phi}},
\end{eqnarray}
\begin{eqnarray}\label{Psi_M Phi_M}
		\Psi(M) = \frac{(M-\sqrt{3 \alpha})^{2}}{2},
		\Phi(M) = \frac{(M+\sqrt{3 \alpha})^{2}}{2}.
\end{eqnarray}

\noindent Although $V(\phi)$ is a function of $\phi$, $M$, $\beta_{1}$, $\mu$ and $\alpha$, but for simplicity, we can omit any argument from $V(M,\phi,\beta_{1},\mu,\alpha)$ when no particular emphasis is put upon it. For example, when we write $V(\phi)=V(M,\phi)$, it is meant that the arguments $\beta_{1}$, $\mu$ and $\alpha$ assume fixed values in their physically admissible range and for the fixed values of $\beta_{1}$, $\mu$ and $\alpha$, nature of solitary structures as obtained from the energy integral (\ref{energy int}) depends only on $M$.

\noindent One can take any equation dynamically equivalent to the energy integral (\ref{energy_integral}) to discuss the qualitative behavior of solitary structures. As the energy integral (\ref{energy_integral}) is dynamically equivalent to the energy integral (\ref{energy int}), we take the energy integral (\ref{energy int}) to discuss the qualitative behavior of solitary structures of the present plasma system. The discussions regarding the existence of solitary structures as given earlier for the energy integral (\ref{energy_integral}) are also true for the energy integral (\ref{energy int}). Now, the discussions regarding the existence of solitary waves and double layers are valid if $\phi = 0$ is an unstable position of equilibrium of a particle of unit mass associated with the energy integral (\ref{energy int}) under the action of the force field $-V'(\phi)$, i.e., if $V''(0) < 0$ along with $V(0) = V'(0) = 0$. In other words, $\phi=0$ can be made an unstable position of equilibrium if the potential energy of the said particle attains its maximum value at $\phi=0$. Now, the condition $V''(0)<0$ gives a lower bound $M_{c}$ of $M$, i.e., $V''(0)<0\Leftrightarrow M>M_{c}$, $V''(0)>0\Leftrightarrow M<M_{c}$, and $V''(0)=0\Leftrightarrow M=M_{c}\Leftrightarrow V''(M_{c},0)=0$. This $M_{c}$ is, in general, a function of the parameters involved in the system, or a constant. Therefore, if $M<M_{c}$, the potential energy of the said particle attains its minimum value at $\phi=0$, and consequently, $\phi=0$ is the position of stable equilibrium of the particle, and in this case, it is impossible to make any oscillation of the particle even when it is slightly displaced from its position of stable equilibrium, and consequently there is no question of existence of solitary waves or double layers for $M<M_{c}$. In other words, for the position of unstable equilibrium of the particle at $\phi=0$, i.e., for $V''(0)<0(\Leftrightarrow M>M_{c})$, the function $V(\phi)$ must be convex within a neighborhood of $\phi=0$ and in this case both type of solitary waves (negative or positive potential) may exist if other conditions are fulfilled. Now suppose that $V''(M_{c},0)=0$ and also $V'''(M_{c},0)=0$, then if $V '''' (M_{c},0)<0$, the potential energy of the said particle attains its maximum value at $\phi=0$ and consequently, $\phi=0$ is the position of unstable equilibrium. On the other hand if $V''(M_{c},0)=0$, $V'''(M_{c},0)=0$, and $V '''' (M_{c},0)>0$, the potential energy of the said particle attains its minimum value at $\phi=0$ and consequently, $\phi=0$ is the position of stable equilibrium of the particle and in this case there is no question of existence of solitary wave solution as well as double layer solution of the energy integral (\ref{energy int}). But if $V'''(M_{c},0)\neq 0$ along with $V(M_{c},0)=V'(M_{c},0)=V''(M_{c},0)=0$, then without going through the complete analytical investigation, it is difficult to predict the existence of solitary wave and/or double layer solution of the energy integral (\ref{energy int}) at $M=M_{c}$. In this situation, i.e., when $V'''(M_{c},0)\neq 0$ along with $V(M_{c},0)=V'(M_{c},0)=V''(M_{c},0)=0$, to continue the physical interpretation for the existence of solitary wave and/or double layer solution of the energy integral (\ref{energy int}) at $M=M_{c}$, \citet{das12} have proved the following important results to confirm the existence of solitary structures at $M=M_{c}$.

If $V(M,0)=V'(M,0)=V''(M_{c},0)=0$, $V'''(M_{c},0)<0$ ($V'''(M_{c},0)>0$), $\partial V/\partial M < 0$ for all $M >0$ and for all $\phi >0$ ($\phi  <0$), the main analytical results for the existence of solitary wave and double layer solution of the energy integral at $M=M_{c}$ are as follows.\\
\textbf{Result-1:} If there exists at least one value $M_{0}$ of $M$ such that the system supports PPSWs (NPSWs) for all  $M_{c}<M< M_{0}$, then there exist either a PPSW (NPSW) or a PPDL (NPDL) at $M=M_{c}$.\\
\textbf{Result-2:} If the system supports only NPSWs (PPSWs) for $M>M_{c}$, then there does not exist PPSW (NPSW) at $M=M_{c}$.\\
\textbf{Result-3:} It is not possible to have coexistence of both positive and negative potential solitary structures at $M=M_{c}$.

Apart from the above results, \citet{das12} have mentioned that the PPDL (NPDL) solution at $M=M_{c}$ is possible only when there exists a PPDL (NPDL) solution in any right neighborhood of $M_{c}$, i.e., PPDL (NPDL) solution at $M=M_{c}$ is possible only when the curve $M=M_{D}$ tends to intersect the curve $M=M_{c}$ at some point in the solution space (or the compositional parameter space showing the nature of existence of the different solitary structures) of the energy integral, where each point of the curve $M=M_{D}$ corresponds to a PPDL (NPDL) solution of the energy integral whenever $M_{D}>M_{c}$.

In the present paper our aim is to study the Dust Ion Acoustic (DIA) solitary structures at $M = M_{c}(=M_{s})$, in an unmagnetized nonthermal plasma consisting of negatively charged dust grains, adiabatic warm ions and nonthermal electrons, where $ M_{c}$ is the smallest possible value of the Mach number $M$, i.e., in general, the solitary structures start to exist for $M>M_{c}$. Dust Ion Acoustic (DIA) solitary structures for $M > M_{c}$ in the above mentioned plasma have been studied extensively by \citet{das12}.

\section{\label{sec:theory}Existence and Polarity of solitons at $M=M_{c}$}
For the present problem, it can be easily checked that $V(M,0)=V'(M,0)=0$ for any value of $M$ and also for any values of the parameters $\beta_{1}$, $\mu$ and $\alpha$, whereas the condition $V''(M,0)<0$ gives $M>M_{c}$, where
\begin{eqnarray}\label{Mc nonthermal}
    M_{c}^{2} = 3 \alpha+\frac{1}{\mu (1-\beta_{1})}.
\end{eqnarray}
From (\ref{Mc nonthermal}), we see that for $M_{c}$ to be real and positive, we must have $\mu>0$ and $0\leq \beta_{1}<1$. As the effective range of $\beta_{1}$ is $0\leq \beta_{1} \leq \beta_{1T}$, where $\beta_{1T} = 4/7\approx 0.571429 \approx 0.6$, $M_{c}$ is well-defined as a real positive quantity for all $0<\mu\leq \mu_{T}$ and $0\leq \beta_{1} \leq \beta_{1T}$.

From the discussions of the section \S \ref{sec:energy_int} we have seen that the existence and polarity of solitons at $M=M_{c}$ depend on following facts:
\begin{itemize}
  \item the sign of the term $\frac{\partial V}{\partial M}$,
  \item the sign of the term $V'''(M_{c}, 0)$,
  \item the existence and polarity of solitons in a right neighbourhood of $M=M_{c}$, i.e., the existence and polarity of solitons for $M_{c}<M<M_{c}+\epsilon$, where $\epsilon >0$ be any real.
\end{itemize}
For the present problem the functions, $\frac{\partial V}{\partial M}$ and $V'''(M_{c}, 0)$  are respectively given by
\begin{eqnarray}\label{delvdelM DIA}
\frac{\partial V}{\partial M} = -M\bigg(\sqrt{n_{i}}-\frac{1}{\sqrt{n_{i}}}\bigg)^{2},
\end{eqnarray}
\begin{equation}\label{v3 at Mc non DIA}
    V'''(M_{c}, 0) = \mu \{12 \alpha \beta_{n}^{3}\mu^{2} + 3 \beta_{n}^{2} \mu -1\},
\end{equation}
where
\begin{equation}\label{betan}
    \beta_{n}=1-\beta_{1}.
\end{equation}
From (\ref{delvdelM DIA}), we see that $\partial V/\partial M<0$ for all $M>0$ and for all $\phi\neq 0$, and consequently, for all $M>0$, the condition $\partial V/\partial M<0$ is satisfied for all $\phi > 0$ as well as $\phi < 0$.

Now we can write (\ref{v3 at Mc non DIA}) in the following convenient form:
\begin{equation}\label{modified v3 at Mc non}
    V'''(M_{c}, 0) = \left\{\begin{array}{lcl}
                       12 \alpha \beta_{n}^{3}\mu (\mu+\mu_{*})(\mu-\mu_{c}) & \mbox{when} & \alpha \neq 0,\\
                       3 \beta_{n}^{2}\mu (\mu-\mu_{c}) & \mbox{when} & \alpha = 0,
                     \end{array}\right.
\end{equation}
where
\begin{equation}\label{mu2c}
    \mu_{*}=\mu_{*}(\alpha, \beta_{1})=\frac{1+b(\alpha, \beta_{1})}{8 \alpha \beta_{n}},
\end{equation}
\begin{equation}\label{mu1c}
    \mu_{c}=\mu_{c}(\alpha, \beta_{1})=\frac{2}{3\beta_{n}^{2}[1+b(\alpha, \beta_{1})]},
\end{equation}
\begin{equation}\label{b}
    b=b(\alpha, \beta_{1})=\sqrt{1+\frac{16 \alpha}{3 \beta_{n}}}.
\end{equation}
With the help of simple algebra we have the following observations from equations (\ref{modified v3 at Mc non}) - (\ref{b}):
\begin{itemize}
  \item $\mu_{*}$ is well defined as a strictly positive real number for all $0 < \beta_{n}\leq 1$ and for all $0 < \alpha \leq 1$
  \item $\mu_{c}$ is well defined as a strictly positive real number for all $0 < \beta_{n}\leq 1$ and for all $0 \leq \alpha \leq 1$
  \item the sign of $V'''(M_{c}, 0)$ depends only on the factor $\mu-\mu_{c}$
  \item $V'''(M_{c}, 0)<0$ or $V'''(M_{c}, 0)>0$ according to whether $\mu<\mu_{c}$ or $\mu>\mu_{c}$
  \item if $\mu_{c}\geq 1$, $V'''(M_{c}, 0)<0$ as $0<\mu \leq \mu_{T}<1$
  \item if $0<\mu_{c}<\mu_{T}<1$ then $V'''(M_{c}, 0)<0$ or $V'''(M_{c}, 0)>0$ according to whether $\mu<\mu_{c}$ or $\mu>\mu_{c}$
  \item When $\mu=\mu_{c}$ along with $0<\mu_{c}<\mu_{T}<1$,  $V'''(M_{c}, 0)=0$ and in this case, for the existence of solitary structure at $M=M_{c}$, it is necessary that $V''''(M_{c}, 0)<0$. When $\alpha \neq 0$, with the help of simple algebra, we get the following expression of $V''''(M_{c}, 0)$ at $\mu=\mu_{c}$
  \begin{equation}\label{v4Mc alpha neq zero}
    V''''(M_{c}, 0) = \frac{2[(3 \beta_{n}-2)^{2}(b+1)^{2}+5(b^2-1)+4b]}{9 \beta_{n}^{3}(b+1)^{3}},
  \end{equation}
  whereas for $\alpha = 0$, the expression of $V''''(M_{c}, 0)$ at $\mu=\mu_{c}$ is given by
  \begin{equation}\label{v4Mc alpha eq zero}
    V''''(M_{c}, 0) = \frac{(3 \beta_{n}-2)^{2}+1}{9 \beta_{n}^{3}}.
  \end{equation}
  \item From equation (\ref{v4Mc alpha neq zero}) and equation (\ref{v4Mc alpha eq zero}), one can check that for any admissible values of parameters of the system $V''''(M_{c}, 0)>0$ when $\mu=\mu_{c}$, i.e., when $V'''( M_{c}, 0)=0$. Therefore, there does not exist any solitary wave solution of the energy integral (\ref{energy int}) at $M=M_{c}$ when $V'''(M_{c}, 0)=0$.
  \item Again, differentiating (\ref{modified v3 at Mc non}) with respect to $\beta_{1}$, we get \begin{equation}\label{diff wrt beta1 v3 at Mc non}
        \frac{\partial}{\partial \beta_{1}}(V'''(M_{c}, 0)) = \left\{\begin{array}{lcl}
                       -36 \alpha \mu^{3} \beta_{n}(\beta_{n}+\frac{1}{6 \alpha \mu}) & \mbox{when} & \alpha \neq 0,\\
                       -6 \mu^{2} \beta_{n} & \mbox{when} & \alpha = 0.
                       \end{array}\right.
      \end{equation}
      Using the simple algebra, it is easy to check that if $\mu > \mu_{c}(\alpha,0)$, $V'''(M_{c}, 0)$ strictly decreases with increasing $\beta_{1}$ for all $0 \leq \beta_{1} <1$ starting from a positive value and ending with a negative value, whereas if $\mu < \mu_{c}(\alpha,0)$, $V'''(M_{c}, 0)$ strictly decreases with increasing $\beta_{1}$ for all $0 \leq \beta_{1} <1$ starting from a negative value and ending with a negative value. So, $V'''(M_{c}, 0)$ intersects the axis of $\beta_{1}$ at $\beta_{1}=\beta_{c}(0 \leq \beta_{c}<1)$ only if $\mu > \mu_{c}(\alpha,0)$ and consequently, we have the following conclusions regarding the sign of $V'''(M_{c}, 0)$.
      \begin{itemize}
      \item If $\mu>\mu_{c}(\alpha, 0)$ and $\beta_{c}<\beta_{1T}$, we get $V ''' (M_{c}, 0) > 0$ for all $0\leq \beta_{1}<\beta_{c}$, whereas $V ''' (M_{c}, 0) < 0$ for all $\beta_{c}<\beta_{1}\leq \beta_{1T}$ and $V ''' (M_{c}, 0) = 0$ at $\beta_{1}=\beta_{c}$.
      \item If $\mu>\mu_{c}(\alpha, 0)$ and $\beta_{c}>\beta_{1T}$, we get $V ''' (M_{c}, 0) > 0$ for all $0\leq \beta_{1}\leq\beta_{1T}$.
      \item If $\mu<\mu_{c}(\alpha, 0)$, we get $V ''' (M_{c}, 0) < 0$ for all $0\leq \beta_{1}\leq\beta_{1T}$.
\end{itemize}
    \item From the above discussions, it is interesting to note that when $\beta_{1} = \beta_{c}$ then $\mu = \mu_{c}$, since for $\beta_{1} = \beta_{c}$, we have $V ''' (M_{c}, 0) = 0$ and the equation $V ''' (M_{c}, 0) = 0$ has only one solution $\mu = \mu_{c}$ for $\mu$ within the admissible range of $\mu$. On the other hand when $\mu = \mu_{c}$ then $\beta_{1} = \beta_{c}$ , since for $\mu = \mu_{c}$, we have $V ''' (M_{c}, 0) = 0$ and the equation $V ''' (M_{c}, 0) = 0$ has only one solution $\beta_{1} = \beta_{c}$ for $\beta_{1}$ within the admissible range of $\beta_{1}$. Therefore, we can conclude that when $\mu = \mu_{c}$ then $\beta_{1} = \beta_{c}$ and conversely, when $\beta_{1} = \beta_{c}$ then $\mu = \mu_{c}$ and consequently, at $\beta_{1} = \beta_{c}$, we have $V '''' (M_{c}, 0) > 0$, i.e., there does not exist any solitary wave solution and/or double layer solution of the energy integral at $M=M_{c}$ when $\beta_{1} = \beta_{c}$($\Leftrightarrow$ $\mu = \mu_{c}$).
\end{itemize}

Now we  are in a position to discuss the existence and polarity of the solitary structures of the energy integral (\ref{energy int}) at $M=M_{c}$ with the help of the qualitatively different solution spaces i.e., the compositional parameter spaces showing the nature of existence of the different solitary structures for $M>M_{c}$ and the theoretical discussions as given earlier in this section and in the introduction of the present paper. The qualitatively different solution spaces for $M>M_{c}$ are necessary to discuss the existence and polarity of the solitary structures at $M=M_{c}$ because in the introduction of the present paper we have seen that one can apply \textbf{Result-1 or Result-2} if and only if we have a definite idea regarding the existence and the polarity of the solitary structures in the right neighbourhood of $M=M_{c}$. To introduce the solution space it is necessary to define the cut off values of the parameters involved in the system. Although these cut off values of the parameters have already been defined in the paper of \citet{das12}, but for easy readability of the present paper we have recapitulated those cut off values of the parameters to clarify the four qualitatively different solution spaces. The analytical theory for the construction of the solution spaces have been given in different articles of \citet{das09,das10,das12}.

\begin{figure}
\begin{center}
  \includegraphics{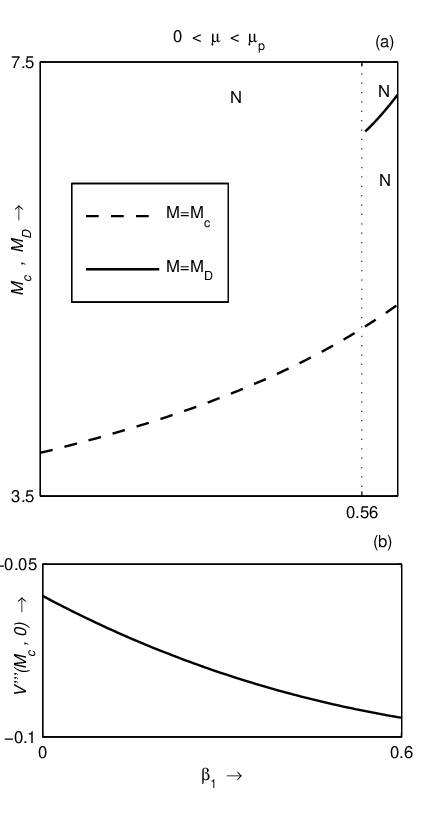}
  \caption{\label{fig:sol space1} (a) Solution space, i.e., the compositional parameter space showing the existence of different solitary structures with respect to $\beta_{1}$ for $\alpha=0.9$ and for a particular value of $\mu$ lying within the interval $0 < \mu < \mu_{p}$. Here N stands for the occurrence of NPSW and NPSW starts to exist for $M>M_{c}$ except at the point on the curve $M=M_{D}$.  At any point on the curve $M=M_{D}$, one can always find a NPDL solution. This solution space has been drawn for $\alpha=0.9$ and $\mu=0.1$. For $\alpha = 0.9$, we have found $\mu_{p}  \approx 0.14$ with $\beta_{1c} \approx 0.56$.  (b) $V'''(M_{c} , 0)$ is plotted against $\beta_{1}$ for the same values of $\alpha$ and $\mu$.}
\end{center}
\end{figure}

\begin{figure}
\begin{center}
  \includegraphics{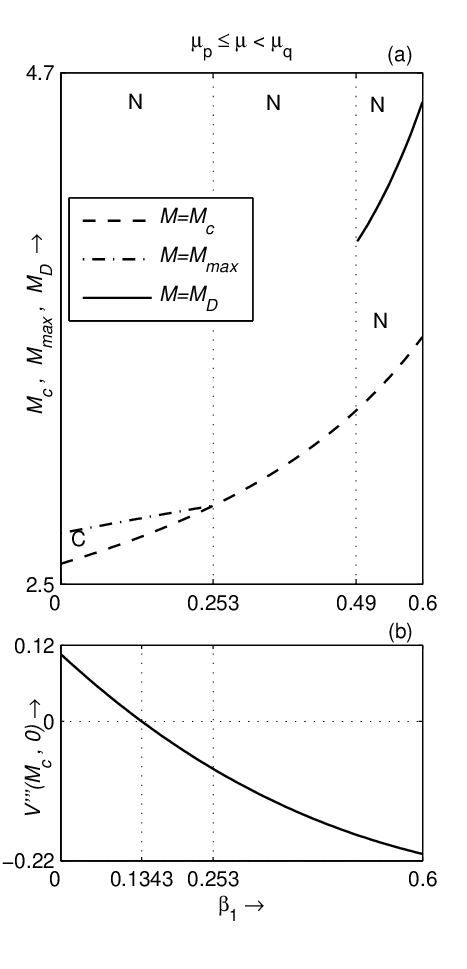}
  \caption{\label{fig:sol space2} (a) Solution space with respect to $\beta_{1}$ for $\alpha=0.9$ and for a particular value of $\mu$ lying within the interval $\mu_{p} \leq \mu < \mu_{q}$. Here N stands for the occurrence of NPSW and NPSW starts to exist for $M>M_{c}$ except at the point on the curve $M=M_{D}$.  At any point on the curve $M=M_{D}$, one can always find a NPDL solution. Here C stands for the simultaneous occurrence of both NPSW and PPSW. This solution space has been drawn for $\alpha=0.9$ and $\mu=0.25$. For $\alpha = 0.9$, we have found $\mu_{p}  \approx 0.14$ and $\mu_{q}  \approx 0.36$ with $\beta_{1a} \approx 0.253$ $\beta_{1c} \approx 0.49$.  (b) $V'''(M_{c} , 0)$ is plotted against $\beta_{1}$ for the same values of $\alpha$ and $\mu$. Note that $\beta_{c}=0.134$ and $0<\beta_{c}<\beta_{1a}.$}
\end{center}
\end{figure}

\begin{figure}
\begin{center}
  \includegraphics{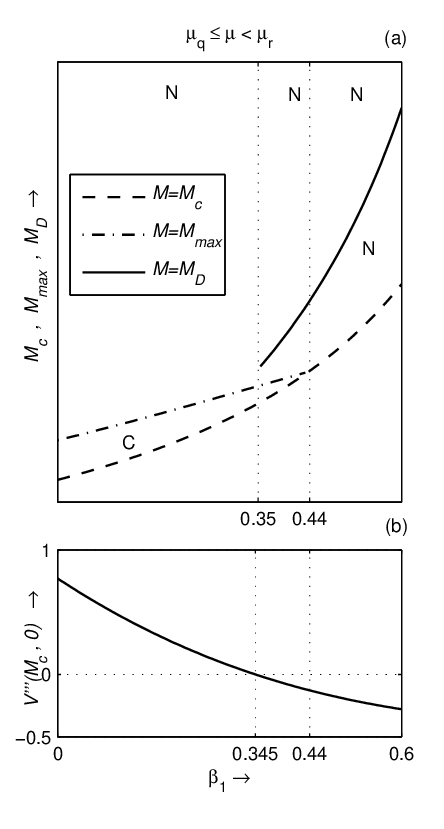}
  \caption{\label{fig:sol space3} (a) Solution space with respect to $\beta_{1}$ for $\alpha=0.9$ and for a particular value of $\mu$ lying within the interval $\mu_{q} \leq \mu < \mu_{r}$. Here N stands for the occurrence of NPSW and NPSW starts to exist for $M>M_{c}$ except at the point on the curve $M=M_{D}$.  At any point on the curve $M=M_{D}$, one can always find a NPDL solution. Here C stands for the simultaneous occurrence of both NPSW and PPSW. This solution space has been drawn for $\alpha=0.9$ and $\mu=0.4$. For $\alpha = 0.9$, we have found $\mu_{q}  \approx 0.36$ and $\mu_{r}  \approx 0.44$ with $\beta_{1a} \approx 0.437$ and $\beta_{1c} \approx 0.352$.  (b) $V'''(M_{c} , 0)$ is plotted against $\beta_{1}$ for the same values of $\alpha$ and $\mu$. Note that $\beta_{c}=0.345$ and $0<\beta_{c}<\beta_{1a}.$}
\end{center}
\end{figure}

\begin{figure}
\begin{center}
  \includegraphics{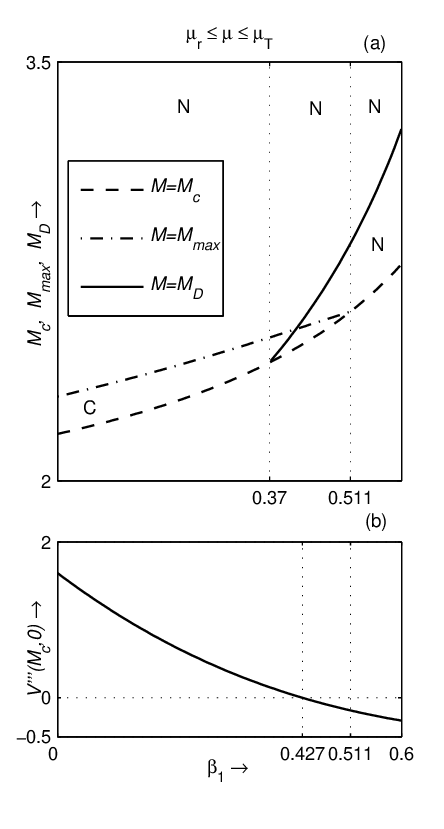}
  \caption{\label{fig:sol space4} Solution space with respect to $\beta_{1}$ for $\alpha=0.9$ and for a particular value of $\mu$ lying within the interval $\mu_{r} \leq \mu \leq \mu_{T}$. Here N stands for the occurrence of NPSW and NPSW starts to exist for $M>M_{c}$ except at the point on the curve $M=M_{D}$.  At any point on the curve $M=M_{D}$, one can always find a NPDL solution. Here C stands for the simultaneous occurrence of both NPSW and PPSW. This solution space has been drawn for $\alpha=0.9$ and $\mu=0.5$. For $\alpha = 0.9$, we have found $\mu_{r}  \approx 0.44$ with $\beta_{1a} \approx 0.511$ and $\beta_{1c} \approx 0.37$.  (b) $V'''(M_{c} , 0)$ is plotted against $\beta_{1}$ for the same values of $\alpha$ and $\mu$. Note that $\beta_{c}=0.427$ and $0<\beta_{c}<\beta_{1a}.$}
\end{center}
\end{figure}

\begin{figure}
\begin{center}
  \includegraphics{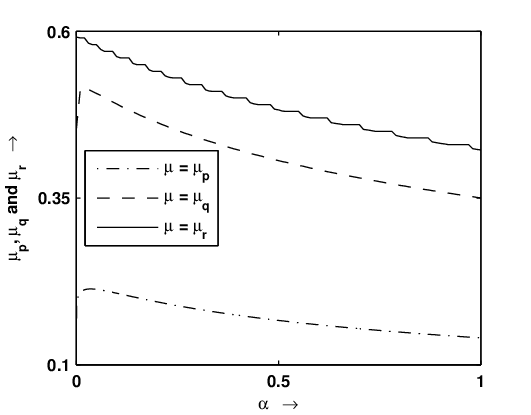}
  \caption{\label{fig:mu_pqr} $\mu_{p}$, $\mu_{q}$ and $\mu_{r}$ are plotted against $\alpha$. Note that $\mu_{p} < \mu_{q} < \mu_{r}$ for any value of $\alpha$.}
\end{center}
\end{figure}

Figure \ref{fig:sol space1}(a) - Figure \ref{fig:sol space4}(a) are the different compositional parameter spaces with respect to $\beta_{1}$ showing nature of solitary structures and all these figures are aimed to show the solution spaces of the energy integral (\ref{energy int}) with respect to $\beta_{1}$. To interpret Figure \ref{fig:sol space1}(a) - Figure \ref{fig:sol space4}(a), we have made a general description as follows: solitary structures start to exist just above the lower curve $M = M_{c}$. For any admissible range of the parameters there always exists at least one $M>M_{c}$ such that NPSW exists thereat. $M_{max}$ is the upper bound of $M$ for the existence of PPSWs, i.e., there does not exist any PPSW if $M>M_{max}$. More explicitly, if we pick a $\beta_{1}$ and goes vertically upwards, then all intermediate $M$ bounded by $M = M_{c}$ and $M=M_{max}$ would give PPSWs. The curve $M=M_{max}$ also restrict the coexistence of both NPSWs and PPSWs, however the curve $M=M_{max}$ is unable to restrict the occurrence of all NPSWs of the present system, i.e., there exists NPSW for all $M>M_{max}$. At any point on the curve $M=M_{D}$ there exists an NPDL solution. But this NPDL solution is unable to restrict the occurrence of all NPSWs of the present system. As a result, we get two different types of NPSWs separated by the NPDL solution, in which occurrence of first type of NPSW is restricted by $M_{c}<M<M_{D}$, whereas the second type NPSW exists for all $M>M_{D}$. We have also observed a finite jump between the amplitudes of NPSWs at $M=M_{D}-\epsilon_{1}$ and at $M=M_{D}+\epsilon_{2}$, where $0<\epsilon_{1}<M_{D}-M_{c}$ and $\epsilon_{2}>0$, i.e., there is a finite jump in amplitudes of the NPSWs above and below the curve $M=M_{D}$. Now we want to define the cut off values of $\mu$ and $\beta_{1}$, which are responsible to delimit the solution space.
\begin{description}
  \item[$\mathbf{\beta_{1c}}$ : ] $\beta_{1c}$ is a cut-off value of $\beta_{1}$ such that NPDL starts to exist whenever $\beta_{1} > \beta_{1c}$ for any value of $\mu$ lies within the interval $0 < \mu \leq \mu_{T} < 1$, $\mu_{T}$ is a physically admissible upper bound of $\mu$, i.e., $\beta_{1} = \beta_{1c}$ is the lower bound of $\beta_{1}$ for the existence of NPDL solution. Thus, $\beta_{1c}$ is the minimum proportion of fast energetic electrons such that maximum potential difference occurs at $\beta_{1} = \beta_{1c}+\epsilon$ for any $\epsilon >0$ and for fixed values of other parameters of the system. Therefore, for $\beta_{1} > \beta_{1c}$, there exists a value $M_{D}$ ($> M_{c}$) of $M$ such that one can get an NPDL as a solution of the energy integral (\ref{energy int}) at $M = M_{D}$. In other words, for $\beta_{1} > \beta_{1c}$, there exists a sequence of NPSWs of increasing amplitude restricted by $M_{c} < M < M_{D}$ such that this sequence of NPSWs converges to NPDL at $M = M_{D}$, i.e., $M = M_{D}$ can be regarded as an upper bound of the Mach number for the existence of at least one sequence (class) of NPSWs.
	\item[$\mathbf{\mu_{p}}$ : ] $\mu_{p}$ is a cut of value of $\mu$ such that $M_{max}$ does not exist for any admissible value of $\beta_{1}$ if $\mu$ lies within the interval $0<\mu<\mu_{p}$, i.e., if $\mu \geq \mu_{p}$, there exists a value $\beta_{1}^{*}$ of $\beta_{1}$ such that $M_{max}$ exists at $\beta_{1}=\beta_{1}^{*}$, moreover, if $\beta_{1}^{*}>0$, then $M_{max}$ exists for all $\beta_{1}$ lies within the interval $0\leq \beta_{1}<\beta_{1}^{*}$.
	\item[$\mathbf{\beta_{1a}}$ : ]$\beta_{1a}$ is a cut-off value of $\beta_{1}$ such that $M_{max}$ exists for all $0\leq \beta_{1}\leq \beta_{1a}$ whenever $0 < \mu \leq \mu_{T} < 1$. Consequently, $\beta_{1} = \beta_{1a}$ is the upper bound of $\beta_{1}$ for the existence of PPSW, i.e., there does not exist any PPSW for $\beta_{1} > \beta_{1a}$
\end{description}
Now, if $\beta_{1c} > \beta_{1a}$, then there exists an interval $\beta_{1a}<\beta_{1}<\beta_{1c}$ in which neither $M_{D}$ nor $M_{max}$ exist and consequently, we can define cut-off values $\mu_{q}$ and $\mu_{r}$ of $\mu$ as follows:
\begin{description}
  \item[$\mu_{q}$ : ]$\mu_{q}$ is another cut-off value of $\mu$ such that for all $\mu_{p}\leq \mu < \mu_{q}$, neither $M_{max}$ nor $M_{D}$ exist whenever $\beta_{1a}<\beta_{1}< \beta_{1c}$, i.e., for all $\mu_{p}\leq \mu < \mu_{q}$ and for all $\beta_{1a}<\beta_{1}< \beta_{1c}$ only NPSWs exist for all $M>M_{c}$.
  \item[$\mu_{r}$ : ]$\mu_{r}$ is another cut-off value of $\mu$ such that for all $\mu_{r}\leq \mu\leq \mu_{T}$, the curve $M=M_{D}$ tends to intersect the curve $M=M_{c}$ at the point $\beta_{1} = \beta_{1c}$.
\end{description}
From the definition of $\mu_{p}$, $\mu_{q}$ and $\mu_{r}$, we can numerically find the values of $\mu_{p}$, $\mu_{q}$ and $\mu_{r}$ for any value of $\alpha$. The numerical solution is shown graphically in Figure \ref{fig:mu_pqr}. From this Figure we see that for any value of $\alpha$, we can partition the entire interval of $\mu$ in the following four subintervals: (i) $0<\mu< \mu_{p}$, (ii) $\mu_{p}\leq \mu < \mu_{q}$, (iii) $\mu_{q}\leq \mu < \mu_{r}$ and (iv) $\mu_{r}\leq \mu \leq \mu_{T}$. In these subintervals of $\mu$, we have qualitatively different solution space of the energy integral (\ref{energy int}) with respect to $\beta_{1}$. The different solution spaces have been shown through Figure \ref{fig:sol space1}(a) - Figure \ref{fig:sol space4}(a) for four different subintervals of $\mu$.

Consider the solution space as shown in Figure \ref{fig:sol space1}(a). For this solution space ($0 < \mu < \mu_{p}$). For $0 \leq \beta_{1} \leq \beta_{1T}$, Figure \ref{fig:sol space1}(b) shows that $V'''(M_{c}, 0) < 0$ and Figure \ref{fig:sol space1}(a) shows the existence of NPSWs only in any right neighbourhood of the curve $M=M_{c}$ $\Rightarrow$ $V'''(M_{c}, 0) < 0$ and the existence of NPSWs only in any right neighbourhood of the curve $M=M_{c}$ $\Rightarrow$ there does not exist any solitary structures at $M=M_{c}$, where we have used \textbf{Result-2}. Therefore, for $0 \leq \beta_{1} \leq \beta_{1T}$ and for any admissible value of $\alpha$, there does not exist any solitary structure along the curve $M=M_{c}$ if $0 < \mu < \mu_{p}$.

\begin{figure}
\begin{center}
  \includegraphics{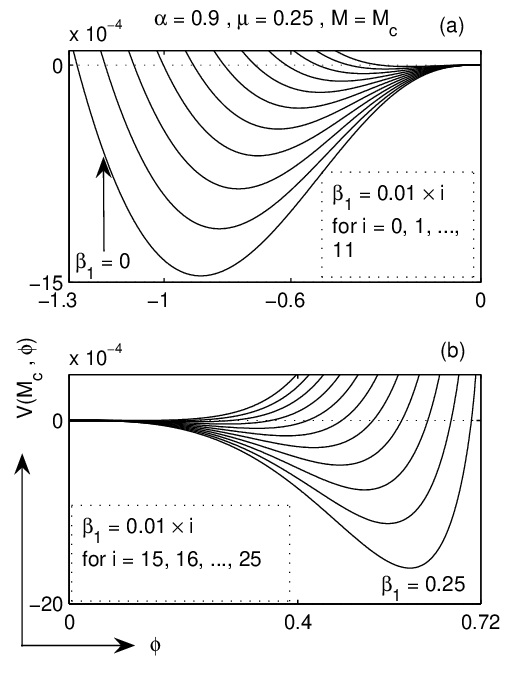}
  \caption{\label{fig:collapse} (a) $V(M_{c},\phi)$ is plotted against $\phi$ for $\alpha=0.9$, $\mu=0.25$ and $\beta_{1} = 0.01 \times i$ where $i=0,1,2,...,11$. (b) $V(M_{c},\phi)$ is plotted against $\phi$ for $\alpha=0.9$, $\mu=0.25$ and $\beta_{1} = 0.01 \times i$ where $i=15,16,...,25$.}
\end{center}
\end{figure}

Consider the solution space as shown in Figure \ref{fig:sol space2}(a). For this solution space ($\mu_{p} \leq \mu < \mu_{q}$).  For $\alpha = 0.9$, $\mu = 0.25$, we find that $\beta_{1c} = 0.49$, $\beta_{1a} = 0.253$, and $\beta_{c} = 0.1343$ which are in agreement of the solution space as shown in Figure \ref{fig:sol space2}(a) and Figure \ref{fig:sol space2}(b), where Figure \ref{fig:sol space2}(b) shows the variation of $V'''(M_{c}, 0)$ against $\beta_{1}$.  Therefore, NPDL starts to exist if $\beta_{1} \geq \beta_{1c}=0.49$ along the curve $M=M_{D}(>M_{c})$ and NPSWs and PPSWs coexist if $0 \leq \beta_{1} \leq \beta_{1a}$ and $M_{c}<M \leq M_{max}$. With respect to the sign  $V'''(M_{c}, 0)$ and the existence of solitary structures in a right neighbourhood of $M=M_{c}$ (as shown in Figure \ref{fig:sol space2}(a)), we can partition the entire interval of $\beta_{1}$ ($0 \leq \beta_{1} \leq \beta_{1T}$) in the following subintervals and using the \textbf{Result-1} and \textbf{Result-2}, one can draw the following conclusions regarding the existence and the polarity of the solitary structures along the curve $M=M_{c}$:
\begin{itemize}
  \item For $0 \leq \beta_{1} < \beta_{c} (= 0.1343)$, Figure \ref{fig:sol space2}(b) shows that $V'''(M_{c}, 0) > 0$ and Figure \ref{fig:sol space2}(a) shows the coexistence of both NPSWs and PPSWs in a right neighbourhood of the curve $M=M_{c}$ $\Rightarrow$ $V'''(M_{c}, 0) > 0$ and the existence of NPSWs in a right neighbourhood of the curve $M=M_{c}$ $\Rightarrow$ the existence of NPSWs at $M=M_{c}$, where we have used \textbf{Result-1}. Therefore, for $0 \leq \beta_{1} < \beta_{c} = 0.1343$, NPSW exists along the curve $M=M_{c}$.
  \item For $\beta_{1} = \beta_{c}$, from equation (\ref{v4Mc alpha neq zero}) for $\alpha \neq 0$ or from equation (\ref{v4Mc alpha eq zero}) for $\alpha =0$, it is simple to check that $V''''(M_{c}, 0) > 0$ and consequently, there does not exist any solitary structures at $\beta_{1} = \beta_{c}$ when $M=M_{c}$
  \item For $\beta_{c} < \beta_{1} \leq \beta_{1a}$, Figure \ref{fig:sol space2}(b) shows that $V'''(M_{c}, 0) < 0$ and Figure \ref{fig:sol space2}(a) shows the coexistence of both NPSWs and PPSWs in a right neighbourhood of the curve $M=M_{c}$ $\Rightarrow$ $V'''(M_{c}, 0) < 0$ and the existence of PPSWs in a right neighbourhood of the curve $M=M_{c}$ $\Rightarrow$ the existence of PPSWs at $M=M_{c}$, where we have used \textbf{Result-1}. Therefore, for $\beta_{c} < \beta_{1} \leq \beta_{1a}$, PPSW exists along the curve $M=M_{c}$.
  \item For $\beta_{1a} < \beta_{1} \leq \beta_{1T}$, Figure \ref{fig:sol space2}(b) shows that $V'''(M_{c}, 0) < 0$ and Figure \ref{fig:sol space2}(a) shows the existence of NPSWs only in a right neighbourhood of the curve $M=M_{c}$ $\Rightarrow$ $V'''(M_{c}, 0) < 0$ and the existence of NPSWs in a right neighbourhood of the curve $M=M_{c}$ $\Rightarrow$ there does not exist any solitary structures at $M=M_{c}$, where we have used \textbf{Result-2}. . Therefore, for $\beta_{c} < \beta_{1} \leq \beta_{1a}$, there does not exist any solitary structures along the curve $M=M_{c}$.
  \item For this value of $\mu (=0.25)$, the present system does not support any double layer solution at $M=M_{c}$ because from the discussion as given in the introduction of the present paper, we have seen that a system can support a NPDL (PPDL) at $M=M_{c}$ if the curve $M=M_{D}$ for NPDL (PPDL) tends to intersect the curve $M=M_{c}$ at some point of the solution space, i.e., if there exists a double layer solution of same polarity in every right neighbourhood of $M=M_{c}$.
  \item In support of the above conclusions drawn from \textbf{Result-1} and \textbf{Result-2}, we draw Figure \ref{fig:collapse} which shows the variation of $V(\phi)$ against $\phi$. To be more specific, Figure \ref{fig:collapse}(a) shows the variation of $V(\phi)$ against $\phi$ for $\alpha = 0.9$, $\mu = 0.25$ for different values of $\beta_{1}$ starting from $\beta_{1}=0$ ending with $\beta_{1} = 0.11$ at $M=M_{c}$. Note that for the above mentioned values of $\alpha$ and $\mu$ the value of $\beta_{c}$ is equal to 0.1343. From Figure \ref{fig:collapse}(a), we see that for $0 \leq \beta_{1} < \beta_{c} = 0.1343$, NPSW exists along the curve $M=M_{c}$ and the amplitude of NPSW decreases with increasing $\beta_{1}$ when $\beta_{1}$ lying within the interval $0 \leq \beta_{1} < \beta_{c} = 0.1343$, and ultimately NPSW collapses at $ \beta_{1} =\beta_{c}$. On the other hand, Figure \ref{fig:collapse}(b) shows the variation of $V(\phi)$ against $\phi$ for $\alpha = 0.9$, $\mu = 0.25$ for different values of $\beta_{1}$ starting from $\beta_{1}=0.15$ ending with $\beta_{1} = 0.25$ at $M=M_{c}$. Note that for the above mentioned values of $\alpha$ and $\mu$ the value of $\beta_{c}$ is equal to 0.1343 and $\beta_{1a} = 0.253$. From Figure \ref{fig:collapse}(b), we see that for $\beta_{c} < \beta_{1} \leq \beta_{1a} $, PPSW exists along the curve $M=M_{c}$ and the amplitude of PPSW decreases with decreasing $\beta_{1}$ when $\beta_{1}$ lying within the interval $\beta_{c} < \beta_{1} \leq \beta_{1a} $, and ultimately PPSW also collapses at $ \beta_{1} =\beta_{c}$. Therefore Figure \ref{fig:collapse}(a) and Figure \ref{fig:collapse}(b) show that both PPSW and NPSW collapse at $ \beta_{1} =\beta_{c}$ and this is in agreement with statement that there does not exist any solitary structures at $\beta_{1} = \beta_{c}$ when $M=M_{c}$.
\end{itemize}

Exactly by the similar argument as given for the solution space Figure \ref{fig:sol space2}(a) ($\mu_{p} \leq \mu < \mu_{q}$) regarding the existence of solitary structures along the curve $M=M_{c}$, one can study the solution space Figure \ref{fig:sol space3} (a) ($\mu_{q} \leq \mu < \mu_{r}$) to investigate the existence and the polarity of solitary structures along the curve $M=M_{c}$. Using the \textbf{Result-1} and \textbf{Result-2} as mentioned in the introduction of the present paper, we arrive exactly qualitatively similar observations as as given for the solution space Figure \ref{fig:sol space2}(a) ($\mu_{p} \leq \mu < \mu_{q}$) regarding the existence of solitary structures along the curve $M=M_{c}$, the only exceptions are the differences between the numerical values of $\beta_{1c}$, $\beta_{1a}$ and $\beta_{c}$. In fact, here for $\alpha=0.9$, $\mu=0.4$, the values of $\beta_{1c}$, $\beta_{1a}$ and $\beta_{c}$ are 0.35, 0.44 and 0.345 respectively, which are in agreement of the solution space as shown in Figure \ref{fig:sol space3}(a) and the Figure \ref{fig:sol space3}(b).

\begin{figure}
\begin{center}
  \includegraphics{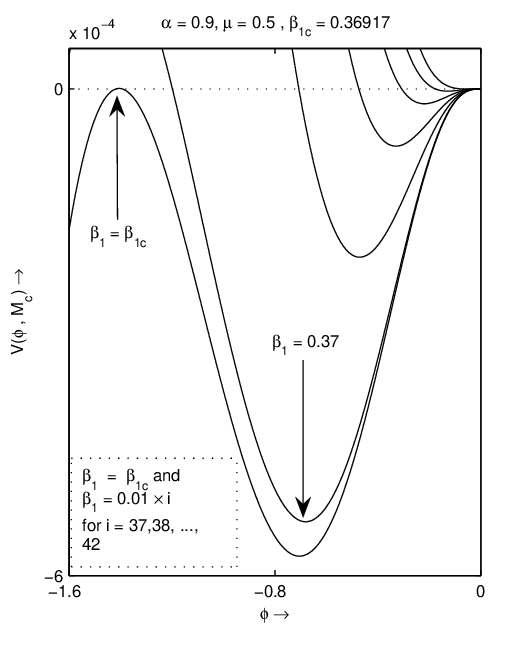}
  \caption{\label{fig:convergence} $V(M_{c},\phi)$ is plotted against $\phi$ for $\alpha=0.9$, $\mu=0.5$ and for each $\beta_{1} = 0.01 \times i$ where $i=37,38,39,40,41,42$ and $\beta_{1}=\beta_{1c}=0.36917$.}
\end{center}
\end{figure}

For the solution space Figure \ref{fig:sol space4}(a) ($\mu_{r} \leq \mu \leq \mu_{T}$), one can investigate the existence and the polarity of solitary structures along the curve $M=M_{c}$ using the exactly similar logic as mentioned for the solution space Figure \ref{fig:sol space2}(a) ($\mu_{p} \leq \mu < \mu_{q}$) regarding the existence of solitary structures along the curve $M=M_{c}$. Using the \textbf{Result-1} and \textbf{Result-2} as mentioned in the introduction of the present paper, we arrive the similar observations as given for the solution space Figure \ref{fig:sol space2}(a) ($\mu_{p} \leq \mu < \mu_{q}$) regarding the existence of solitary structures along the curve $M=M_{c}$, the exceptions are the differences between the numerical values of $\beta_{1c}$, $\beta_{1a}$ and $\beta_{c}$. In fact, here for $\alpha=0.9$, $\mu=0.5$, the values of $\beta_{1c}$, $\beta_{1a}$ and $\beta_{c}$ are 0.37, 0.0.511 and 0.427 respectively, which are in agreement of the solution space as shown in Figure \ref{fig:sol space4}(a) and the Figure \ref{fig:sol space4}(b). But those are not the only exception for this solution space. The one of the main difference for this solution space (Figure \ref{fig:sol space4}(a)) from the other two solution spaces (Figure \ref{fig:sol space2}(a) $\&$ Figure \ref{fig:sol space3}(a)) is the existence of double layer solution at the point $\beta_{1}=\beta_{1c}$ when $M=M_{c}$. The existence of negative potential double layer solution at the point $\beta_{1}=\beta_{1c}$ when $M=M_{c}$ is also in agreement of the solution space as shown in Figure \ref{fig:sol space4}(a). To discuss this case we consider the interval $0 \leq \beta_{1} < \beta_{c}$. From Figure \ref{fig:sol space4}(a) $\&$ Figure \ref{fig:sol space4}(b), we find that
\begin{itemize}
  \item $V'''(M_{c}, 0) > 0$ $\forall$ $0 \leq \beta_{1} < \beta_{c}$ [from Figure \ref{fig:sol space4}(b)]
  \item $0 \leq \beta_{1c} (= 0.37) < \beta_{c} (= 0.427)$ $\Rightarrow$ $V'''(M_{c}, 0) > 0$ at $\beta_{1} \beta_{1c}$
  \item Figure \ref{fig:sol space4}(a) shows the coexistence of both NPSWs and PPSWs in a right neighbourhood of the curve $M=M_{c}$ except the points along the curve $M=M_{D}$
  \item The curve $M=M_{D}$ tends to intersect the curve $M=M_{c}$ at the point $\beta_{1}=\beta_{1c}$ [ i.e., at the point ($\beta_{1}=\beta_{1c}$, $\alpha = 0.9$ and $\mu=0.5$)] of the solution space.
  \item Every right neighbouhood of $M_{c}$ at the point $\beta_{1}=\beta_{1c}$ contains a point of the curve $M=M_{D}$ for $M_{D}>M_{c}$, i.e., there exists a negative potential double layer solution in every right neighbourhood of $M=M_{c}$ at the point $\beta_{1}=\beta_{1c}$ [ i.e., at the point ($\beta_{1}=\beta_{1c}$, $\alpha = 0.9$ and $\mu=0.5$)] of the solution space.
  \item Therefore, $V'''(M_{c}, 0) > 0$ at $\beta_{1} = \beta_{1c}$ and the existence of a negative potential double layer solution in every right neighbourhood of $M=M_{c}$ at the point $\beta_{1}=\beta_{1c}$ $\Rightarrow$ the existence of a negative potential double layer solution in every right neighbourhood of $M=M_{c}$ at the point $\beta_{1}=\beta_{1c}$.
  \item In support of the above conclusion, we draw Figure \ref{fig:dls at beta 1c}. This figure shows the      variation of $V(M_{c},\phi)$ against $\phi$ at the point $\beta_{1}=\beta_{1c}$ for four different values of  $\mu$. This figure definitely shows the existence of negative potential double layer solution at $M=M_{c}$ when $\beta_{1}$ assumes the value $\beta_{1c}$, where $\beta_{1c}$ is a point in the solution space where the curve $M=M_{D}$ tends to intersect the curve $M=M_{c}$ [actually the parametric point point of the solution space is ($\beta_{1c},\mu,\alpha$), but as $\mu$, $\alpha$ both are given and we take $\beta_{1c}$ as the representative of the parametric point ($\beta_{1c},\mu,\alpha$) ].
  \item These double layer solutions are not localized solitary structures. In fact, in the  literature, we have seen that  when  all  the  parameters  involved in  the  system  assume fixed  values  in  their respective physically  admissible  range,  the  amplitude  of  solitary wave  increases  with  increasing  $M$  and  these  solitary waves  end  with  a  double  layer  of  same  polarity  if  it exists.  Here  it  is  important  to  note  that  $M$  is  not  a function of the parameters involved in the system but is restricted by the inequality $M_{c}<M<M_{D}$, where  $M=M_{D}$ corresponds  to  a  double  layer  solution.  So, we cannot compare this case with the case when $M=M_{c}$, since $M_{c}$ is a function of parameters of the system. But the solitons and double layer are not two distinct nonlinear structures even when $M=M_{c}$. Actually, if double layer solution exists, it must be the limiting structure of at least one sequence of solitons of same polarity even when $M=M_{c}$. More specifically, existence  of  double  layer  solution  implies  that  there must exist a sequence of solitary waves of same polarity having monotonically increasing amplitude converging to the double layer solution, i.e., the amplitude of the double  layer  solution  acts  as  an  exact  upper  bound or  least  upper  bound  (lub)  of  the  amplitudes  of  the sequence of solitary waves. Therefore, if the double layer solution exists at $M=M_{c}$ then this double layer solution is  also  a  limiting  structure  of  a  sequence  of solitary waves of same polarity. In support of this conclusion we draw the Figure \ref{fig:convergence}. Figure \ref{fig:convergence} clearly states that negative potential double layer is  a  limiting  structure  of  a  sequence  of negative potential solitary waves even when $M=M_{c}$. In this figure, it is important to note that as $\beta_{1}$ approaches to $\beta_{1c} = 0.36917$ from the right side of $\beta_{1}$ - axis, the negative potential solitary waves approaches to negative potential double layer at $\beta_{1}=\beta_{1c}$ whereas if $\beta_{1}$ approaches to $\beta_{c} = 0.427$ from the left side of $\beta_{1}$ - axis, the negative potential solitary waves collapse at $\beta_{1}=\beta_{c}$
\end{itemize}
The another interesting difference of the solution space (Figure \ref{fig:sol space4}(a)) from the other two solution spaces (Figure \ref{fig:sol space2}(a) $\&$ Figure \ref{fig:sol space3}(a)) is the existence of super solitons in the left neighbourhood of point $\beta_{1}=\beta_{1c}$ when $M=M_{c}$. The existence of negative potential super solitons in the left neighbourhood of point $\beta_{1}=\beta_{1c}$ when $M=M_{c}$ is also in agreement of the solution space as shown in Figure \ref{fig:sol space4}(a). To discuss this case we consider the interval $0 \leq \beta_{1} < \beta_{c}$. From Figure \ref{fig:sol space4}(a) $\&$ Figure \ref{fig:sol space4}(b). For the interval $0 \leq \beta_{1} < \beta_{c}$, we have the following observations.
\begin{itemize}
  \item $0 \leq \beta_{1c} < \beta_{c}$.
  \item Except the point $\beta_{1} = \beta_{1c}$, the negative potential solitary waves exist at $M=M_{c}$ for every $\beta_{1}$ lying within the interval $0 \leq \beta_{1} < \beta_{c}$ and at $\beta_{1} = \beta_{1c}$, we have a negative potential double layer when $M=M_{c}$.
  \item We have two types of negative potential solitary waves in $0 \leq \beta_{1} < \beta_{c}$ along the curve $M=M_{c}$ separated by the negative potential double layer at $\beta_{1} = \beta_{1c}$ and consequently, we have two disjoint intervals $0 \leq \beta_{1} < \beta_{1c}$ and $0 \beta_{1c} < \beta_{1} < \beta_{c}$. At  each point of $0 \leq \beta_{1} < \beta_{1c}$ and $\beta_{1c} < \beta_{1} < \beta_{c}$ we have a negative potential solitary wave along the curve $M=M_{c}$.
  \item In the previous paragraph (Figure \ref{fig:convergence}), we have seen the negative potential solitary waves along the curve $M=M_{c}$ with in the interval $\beta_{1c} < \beta_{1} < \beta_{c}$ converges to the negative potential double layer at $\beta_{1}=\beta_{1c}$ for decreasing $\beta_{1}$ lying within $\beta_{1c} < \beta_{1} < \beta_{c}$.
  \item We know that if  there exists two types of NPSWs (PPSWs) separated by a NPDL (PPDL) then there is a finite jump between the amplitudes of two types of NPSWs only when $\frac{\partial V}{\partial M}<0$ for all $M > 0$ and all $\phi < 0$ ($\phi > 0$)[\citet{das12}]. From this property of \citet{das12}, it is evident that there exists super solitons at each point of $0 \leq \beta_{1} < \beta_{c}$. We draw Figure \ref{fig:jump} in support of this conclusion. This figure clearly shows the existence of super solitons along the curve $M=M_{c}$ for every $\beta_{1}$ lying within $0 \leq \beta_{1} < \beta_{1c}$. This figure also shows the jump type discontinuity in amplitudes between the two types of NPSWs separated by NPDL at $\beta_{1} = \beta_{1c}$ along the curve $M=M_{c}$.
\end{itemize}

The above discussions regarding different solitary structures along the curve $M=M_{c}$ can be summarized through Figure \ref{fig:sol space}. This figure is nothing but the solution space or the compositional parameter space showing the nature of existence of solitary structure along the curve $M=M_{c}$. Figure is self explanatory. We have drawn this solution space for $\alpha=0.9$. We can generalize this solution space if and only if $\mu_{p}<\mu_{c}<\mu_{r}$ for all admissible values of $\alpha$. For this purpose we draw Figure \ref{fig:muprc}. This figure clearly shows that $\mu_{p}<\mu_{c}<\mu_{r}$ for all admissible values of $\alpha$. In the Figure \ref{fig:sol space}, $P$ stands for the existence of PPSW at $M=M_{c}$, $N$ stands for for the existence of NPSW at $M=M_{c}$ and we have double layer solution at $M=M_{c}$ along the curve $\beta_{1}=\beta_{1c}$.  A closer look at the solution space given by Figure. \ref{fig:sol space} reveals the interesting feature near the curve $\beta_{1} = \beta_{1c}$ for $\mu_{r} \leq \mu_{c} \leq \mu_{T}$. In $\mu_{r} \leq \mu_{c} \leq \mu_{T}$, we see that NPSWs exist for both $\beta_{1} < \beta_{1c}$ and $\beta_{1}> \beta_{1c}$ (at least in a right neighbourhood of $\beta_{1} = \beta_{1c}$) and an NPDL exists at $\beta_{1} = \beta_{1c}$, i.e., NPSWs at $M=M_{c}$ is separated by the curve $\beta_{1} = \beta_{1c}$, and consequently, we have two different types of NPSWs - one of which is restricted by the amplitude of NPDL at $M=M_{c}$ but the amplitude of the second type NPSWs at at $M=M_{c}$ increases with decreasing $\beta_{1}$ and finally, attains its maximum value when $\beta_{1}=0$ at $M=M_{c}$, i.e.,  the second type NPSWs at $M=M_{c}$  is restricted by the amplitude of the NPSW for $\beta_{1}=0$ at $M=M_{c}$. So, the characteristic of NPSWs in $0\leq \beta_{1} < \beta_{1c}$ and $\beta_{1c}<\beta_{1}<\beta_{c}$ are different. In fact, the NPSWs in $0\leq \beta_{1} < \beta_{1c}$ are the super solitons at $M=M_{c}$ whenever $\mu_{r} \leq \mu \leq \mu_{T}$. We have seen earlier that the amplitude of this solitary wave at $\beta_{1c}-\epsilon$ ($\epsilon>0$) is much higher than that of $\beta_{1c}+\epsilon_{1}$ ($0<\epsilon_{1}<\beta_{c}-\beta_{1c}$). Thus there is a jump in amplitude of solitary waves and this phenomena has also been observed in some recent works [\cite{das12, Verheest09}] when $M > M_{c}$. Therefore, we can say that there exists two types of NPSWs, where the first type of NPSWs are restricted by $\beta_{1a}<\beta_{1}<\beta_{c}$ and the second type NPSWs exist for all $0\leq \beta_{1} < \beta_{1a}$. We have observed the similar facts for DIA solitary waves for $M>M_{c}$ [\citet{das12}]. Super solitons exist in DIA solitary structures even when $M = M_{c}$.

In Figure \ref{fig:amp Neg beta1c}, variation in amplitude of NPSWs have been shown for values of $\beta_{1}$ lies within $\beta_{1c}< \beta_{1}<\beta_{c}$, whenever $\mu_{r}\leq \mu \leq \mu_{T}$. This figure shows that the potential drop is maximum at $\beta_{1} = \beta_{1c}$ and this is reason behind the occurrence of NPDL solution at $\beta_{1} = \beta_{1c}$. This figure also shows that amplitude of NPSW decreases with increasing $\beta_{1}$ and ultimately, these NPSWs demolished at $\beta_{1} = \beta_{c}$. Figure \ref{fig:dls at beta 1c} verifies the fact that there exist NPDL solution at $\beta_{1} = \beta_{1c}$ for values of $\mu$ in $\mu_{r}\leq \mu \leq \mu_{T}$. This figure also shows that the amplitude of NPDLs at $M=M_{c}$ decreases with increasing $\mu$ in $\mu_{r} \leq \mu \leq \mu_{T}$. Again Figure \ref{fig:sol space}, shows that $\beta_{1c}$ is an increasing function of $\mu$ and hence the amplitude of NPDLs at $M=M_{c}$ decreases with increasing $\beta_{1}$ as well.

\begin{figure}
\begin{center}
  \includegraphics{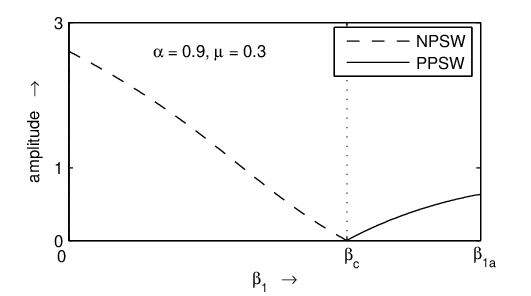}
  \caption{\label{fig:amp both betac} Variation in amplitude of NPSW and PPSW have been shown for values of $\beta_{1}$ lies within $0\leq \beta_{1}<\beta_{c}$ and $\beta_{c}< \beta_{1} \leq \beta_{1a}$. This figure shows that amplitude of NPSW decreases with increasing $\beta_{1}$ and ultimately, these NPSWs demolished at $\beta_{1} = \beta_{c}$, whereas, PPSWs start to occur beyond $\beta_{1} = \beta_{c}$ and amplitude of PPSW increases with $\beta_{1}$ in $\beta_{c}<\beta_{1}\leq \beta_{1a}$ having maximum amplitude at $\beta_{1} = \beta_{1a}$. For $\alpha = 0.9$ and $\mu = 0.3$, we have found $\beta_{c} \approx 0.223$ and $\beta_{1a} \approx 0.33$.}
\end{center}
\end{figure}

In Figure \ref{fig:amp both betac} the variation in amplitude of NPSWs and PPSWs have been shown for values of $\beta_{1}$ lies within $0\leq \beta_{1}<\beta_{c}$ and $\beta_{c}< \beta_{1} \leq \beta_{1a}$ for $\mu_{c} < \mu < \mu_{r}$. This figure shows that amplitude of NPSW decreases with increasing $\beta_{1}$ and ultimately, these NPSWs collapse at $\beta_{1} = \beta_{c}$, whereas PPSWs start to occur beyond $\beta_{1} = \beta_{c}$ and amplitude of PPSW increases with $\beta_{1}$ in $\beta_{c}<\beta_{1}\leq \beta_{1c}$ having maximum amplitude at $\beta_{1} = \beta_{1a}$. Therefore, we can conclude that $\beta_{1} = \beta_{c}$ acts like a sink for NPSWs and source for PPSWs. In fact the same phenomena occurs in $\mu_{r}\leq \mu \leq \mu_{T}$ when $\beta_{1}$ lies within $\beta_{1c}\leq \beta_{1}<\beta_{c}$ and $\beta_{c}< \beta_{1} \leq \beta_{1a}$. But note that we have super solitons for $0 \leq \beta_{1} < \beta_{1c}$ whenever $\mu_{r}\leq \mu \leq \mu_{T}$.

\section{\label{sec:conclusion}Summary $\&$ Discussions}
\begin{figure}
\begin{center}
  \includegraphics{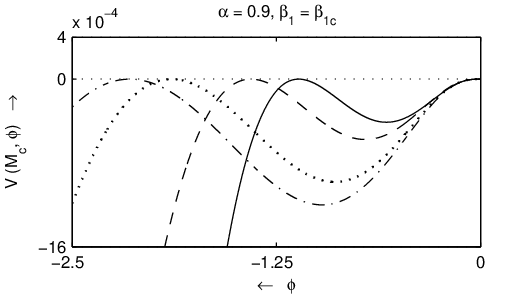}
  \caption{\label{fig:dls at beta 1c} $V(M_{c}, \phi)$ is plotted against $\phi$ and $\beta_{1} = \beta_{1c}$ for four different values of $\mu$, viz., $\mu = 0.44$ (- $\cdot$ -), $\mu = 0.45$ ($\cdots$), $\mu = 0.5$ (- - -) and $\mu = 0.6$ (-----).}
\end{center}
\end{figure}

\begin{figure}
\begin{center}
  \includegraphics{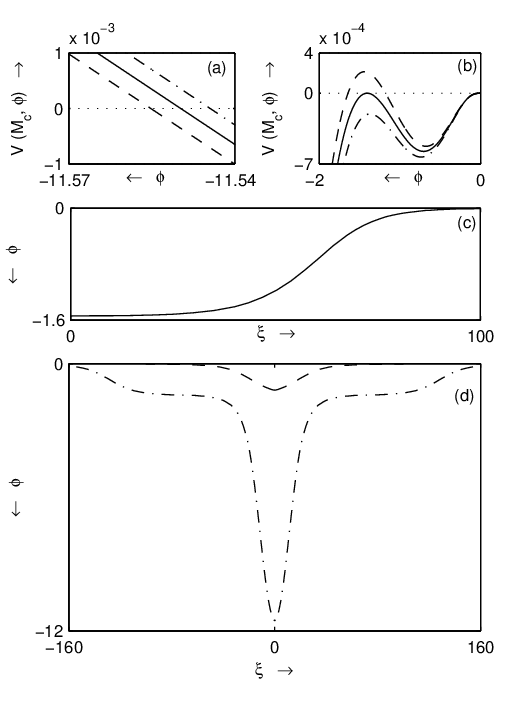}
  \caption{\label{fig:jump} $V(M_{c}, \phi)$ is plotted against $\phi$ for three different values of $\beta_{1}$, viz., $\beta_{1c}$[---], $\beta_{1c}-0.0005$[$- \cdot -$] and $\beta_{1c}+0.0005$[- - -] in (a) and (b). In (b) we have shown the region of $\phi$ from $\phi = -2$ to $\phi = 0$, whereas in (a) a region of $\phi$ from $\phi = -11.57$ to $\phi = -11.54$ has been shown. The NPDL profile corresponding to $\beta_{1} = \beta_{1c}$ has been shown in (c), whereas the profiles of NPSWs corresponding to $\beta_{1}=\beta_{1c}-0.0005$ and $\beta_{1}=\beta_{1c}+0.0005$ have been drawn in (d). Profiles in (d) shown a jump in amplitude of solitary wave by going from $\beta_{1}=\beta_{1c}-0.0005$ to $\beta_{1}=\beta_{1c}+0.0005$. In all these four figures we have used $\alpha = 0.9$, $\mu = 0.5$.}
\end{center}
\end{figure}

\begin{figure}
\begin{center}
  \includegraphics{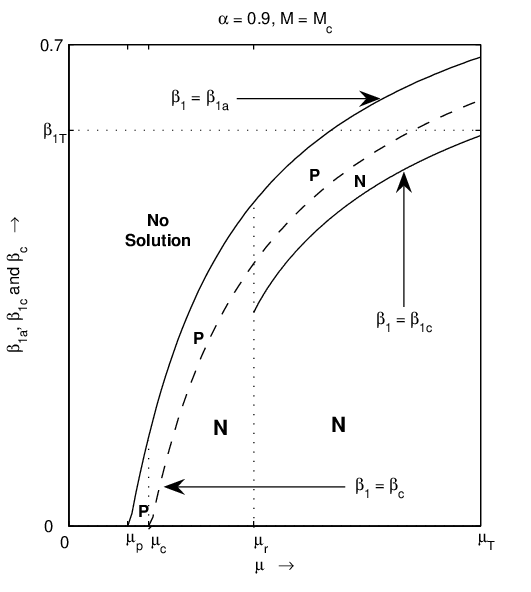}
  \caption{\label{fig:sol space} A graphical presentation of different solitary structures have been given with respect to different subintervals of $\mu$ within the admissible interval of the nonthermal parameter $\beta_{1}$. In this solution space `P' stands for the existence of PPSW and `N' stands for the existence of NPSW. For any given value of $\alpha$, NPSWs exist in $0\leq \beta_{1} < \beta_{c}$ and PPSWs exist in $\beta_{c} < \beta_{1} \leq \beta_{1a}$. At any point on the curve $\beta_{1} = \beta_{1c}$, one can always find a NPDL at $M=M_{c}$. This solution space has been drawn for $\alpha = 0.9$. We have found $\mu_{p}\approx 0.14$, $\mu_{c}\approx 0.2$, and $\mu_{r}\approx 0.44$.}
\end{center}
\end{figure}

\begin{figure}
\begin{center}
  \includegraphics{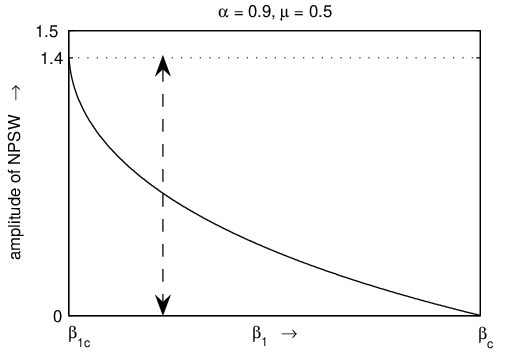}
  \caption{\label{fig:amp Neg beta1c} Variation in amplitude of NPSWs have been shown for values of $\beta_{1}$ lies within $\beta_{1c}< \beta_{1}<\beta_{c}$. This figure shows that amplitude of NPSW decreases with increasing $\beta_{1}$ and ultimately, these NPSWs demolished at $\beta_{1} = \beta_{c}$. For $\alpha = 0.9$ and $\mu = 0.5$, we have found $\beta_{1c} \approx 0.369$ and $\beta_{c} \approx 0.427$.}
\end{center}
\end{figure}

The existence of dust ion acoustic solitary wave and double layer at the lowest possible value of the Mach number $M_{c}$ have been investigated in an unmagnetized nonthermal plasma consisting of negatively charged dust grains, adiabatic positive ions and nonthermal electrons. The general analytical theory of \citet{das12} for the existence of solitary structures at $M=M_{c}$ have been used to investigate the dust ion acoustic solitary wave and double layer at $M=M_{c}$. According to the theorems of \citet{das12}, the solution space for $M>M_{c}$, i.e., the compositional parameter space showing the nature of existence of different solitary structures for $M>M_{c}$ together with the sign of $V'''(M_{c}, \phi)\big|_{\phi=0}$ play the main role to establish the existence of solitary structures at $M=M_{c}$. In fact, if the solitary structure exists at $M=M_{c}$, the sign of $V'''(M_{c}, \phi)\big|_{\phi=0}$ determines the polarity of the solitary structure, specifically, if the solitary structure exists at $M=M_{c}$, then we get a positive or negative potential solitons or double layer according to whether $V'''(M_{c}, \phi)\big|_{\phi=0}<0$ or $V'''(M_{c}, \phi)\big|_{\phi=0}>0$. All the theorems of \citet{das12} have been verified for the present problem. But in the earlier paper, \citet{das12} have not mentioned the analytical theory for the existence of super solitons at $M=M_{c}$. In the present investigation, we have found the existence of super solitons at $M=M_{c}$ for specified range of the parameters involved in the system and this is completely a new observation in the literature of solitons. An analytical approach has been presented to find the polarity of the solitary structures of the present system. Regarding the existence and the polarity of the solitary structures at $M=M_{c}$, we have the following observations:
\begin{enumerate}
  \item For any fixed value of $\alpha$, there exists a value $\mu_{p}$ of $\mu$ such that there do not exist any solitary structures at $M=M_{c}$ for $0<\mu<\mu_{p}$ and for any value of $\beta_{1}$ lying within $0\leq \beta_{1}\leq \beta_{1T}$. In fact, for any value of $\mu$ lies within $0<\mu<\mu_{p}$, there do not exist any non-zero $\phi$ such that $V(M_{c}, \phi) = 0$ for any value of $\beta_{1}$ lying within $0\leq \beta_{1}\leq \beta_{1T}$.
  \item For any value of $\mu$ lies within $\mu_{p}\leq \mu <\mu_{c}$, there exist only PPSWs for all $0 \leq \beta_{1} \leq \beta_{1a}$, where $\mu{c}$ is the root of the equation $V''(M_{c},0)=$ for fixed values of $\alpha$ and $\beta_{1}$ within the specified range.
  \item For any fixed value of $\alpha$, there exists a value $\mu_{r}$ of $\mu$ such that for any value of $\mu$ lying within $\mu_{c} < \mu < \mu_{r}$, NPSWs exist for $0 \leq \beta_{1} <\beta_{c}$ at $M=M_{c}$ and PPSWs exist at $M=M_{c}$ if $\beta_{c} < \beta_{1} \leq \beta_{1a}$.
  \item For any value of $\mu$ lying within $\mu_{r} \leq \mu \leq \mu_{T}$, NPSWs exist in both $0 \leq \beta_{1} <\beta_{1c}$ and $\beta_{1c} < \beta_{1} < \beta_{c}$, whereas PPSWs exist for all $\beta_{c} < \beta_{1} \leq \beta_{1a}$. But for any point on the curve $\beta_{1} = \beta_{1c}$, one can always find a NPDL solution at $M=M_{c}$ when $\mu$ lying within $\mu_{r} \leq \mu \leq \mu_{T}$. These findings have been presented in figure \ref{fig:sol space}. This figure gives the exact scenario regarding the nature of existence of the solitons for the present system at $M=M_{c}$.
  \item A closer look at the solution space given by figure \ref{fig:sol space} reveals an interesting feature near the curve $\beta_{1} = \beta_{1c}$. We have seen the existence of NPSWs at $M=M_{c}$ when $\mu$ lies within $\mu_{r} \leq \mu \leq \mu_{T}$ for all $0 \leq \beta_{1} \leq \beta_{1c}$ and $\beta_{1c} \leq \beta_{1}\leq \beta_{c}$ (i.e., for at least one right neighborhood of $\beta_{1} = \beta_{1c}$) and a NPDL exists at $M=M_{c}$ when $\beta_{1} = \beta_{1c}$. The immediate question is whether the characteristic of NPSWs in $0\leq \beta_{1} < \beta_{1c}$ and $\beta_{1c}<\beta_{1}<\beta_{c}$ are same. In search of the answer, we draw figure \ref{fig:jump}. In figures \ref{fig:jump}(a) and \ref{fig:jump}(b), $V(M_{c}, \phi)$ is plotted against $\phi$ for three different values of $\beta_{1}$, viz., $\beta_{1c}$, $\beta_{1c}-0.0005$ and $\beta_{1c}+0.0005$. Figure \ref{fig:jump}(b) shows that there exists a NPDL at $\beta_{1}=\beta_{1c}$, whereas, for $\beta_{1}=\beta_{1c}+0.0005$, there is a negative potential soliton. However the curve corresponding to $\beta_{1}=\beta_{1c}-0.0005$ has no real root of $\phi$ in the neighborhood of $\phi=\phi_{dl}$, where $\phi_{dl}$ is the amplitude of the NPDL at $\beta_{1}=\beta_{1c}$. We notice from figure (a) that $V(M_{c}, \phi)$ again vanishes far beyond $\phi=\phi_{dl}$ for $\beta_{1}=\beta_{1c}$,  $\beta_{1}=\beta_{1c}+0.0005$, and $\beta_{1}=\beta_{1c}-0.0005$. The curve corresponding to $\beta_{1}=\beta_{1c}-0.0005$ gives a NPSW at $M=M_{c}$ whose amplitude is much greater than the amplitude of NPSW at $\beta_{1}=\beta_{1c}+0.0005$ as well as the amplitude of NPDL at $\beta_{1}=\beta_{1c}$. Thus there is a jump in amplitude of solitary waves and this phenomena has also been observed in some recent works [\citet{Baluku10b,Verheest09,das12}] for the solitary structures when $M>M_{c}$. Therefore, we can say that there exists two types of NPSW at $M=M_{c}$, where the first type solitary wave is restricted by $0\leq \beta_{1} < \beta_{1c}$ and the second type solitary waves exist for all $\beta_{1c}<\beta_{1}<\beta_{c}$. Moreover, at $M=M_{c}, $there is a jump in amplitude between two types of solitary waves corresponding to the nonthermal parameter $\beta_{1}=\beta_{1c}-\epsilon$  and $\beta_{1}=\beta_{1c}+\epsilon$, where $\epsilon$ is a sufficiently small positive quantity. This observation is new in the literature of solitons. In figure \ref{fig:amp Neg beta1c} variation in amplitude of NPSWs have been shown for values of $\beta_{1}$ lies within $\beta_{1c}< \beta_{1}<\beta_{c}$ whenever $\mu_{r}\leq \mu \leq \mu_{T}$. This figure shows that the potential drop is maximum at $\beta_{1} = \beta_{1c}$ and this is reason behind the occurrence of NPDL solution at $\beta_{1} = \beta_{1c}$. This figure also shows that amplitude of NPSW decreases with increasing $\beta_{1}$ and ultimately, these NPSWs will collapse at $\beta_{1} = \beta_{c}$. Figure \ref{fig:dls at beta 1c} verifies the fact that there exist NPDL solution at $\beta_{1} = \beta_{1c}$ for values of $\mu$ in $\mu_{r}\leq \mu \leq \mu_{T}$. This figure also shows that the amplitude of NPDLs at $M=M_{c}$ decreases with increasing $\mu$ in $\mu_{r} \leq \mu \leq \mu_{T}$. Again, figure \ref{fig:sol space}, shows that $\beta_{1c}$ is an increasing function of $\mu$ and hence the amplitude of NPDLs at $M=M_{c}$ decreases with increasing $\beta_{1}$ as well.
  \item It is important to note that the Figure \ref{fig:sol space} not only corresponds to nonthermal distribution electrons but also corresponds to the solution space when the electrons are iso-thermally distributed if we move the solution space along the axis of $\beta_{1}$, i.e., if we put $\beta_{1}=0$. In this case, we can find different intervals of $\mu$ on the basis of the existence and the polarity of solitary structures at $M=M_{c}$ in presence of isothermal electrons. In this case, we have seen that there does not exist any solitary structure at $M=M_{c}$ for $0<\mu<\mu_{p}$. Solitary Structure at $M=M_{c}$ starts to exist if $\mu_{p} \leq \mu \leq \mu_{T}$ except the point $\mu=\mu_{c}$, where $V''(M_{c} , 0)=0$ ( i.e., $\mu=\mu_{c}$ is the root of the equation $V''(M_{c} , 0)=0$, for given value of $\alpha$ and $\beta_{1}=0$).  Only PPSWs can be found for all $\mu$ lying within the interval $\mu_{p}\leq \mu <\mu_{c}$, whereas for $\mu$ lying within $\mu_{c} < \mu \leq \mu_{T}$, there exist only NPSWs. In this case also both NPSWs and PPSWs collapse at $\mu=\mu_{c}$. In support of the above conclusions drawn from the solution space (Figure \ref{fig:sol space}), we draw figure \ref{fig:verify iso}. In this figure, $V(M_{c}, \phi)$ is plotted against $\phi$ for different values of $\mu$. In figure \ref{fig:verify iso}(a), the values of $\mu$ have been taken from $\mu_{c} < \mu \leq \mu_{T}$ whereas in figure \ref{fig:verify iso}(b), the values of $\mu$ have been taken from $\mu_{p}\leq \mu <\mu_{c}$. Figures \ref{fig:verify iso}(a) and \ref{fig:verify iso}(b) verifies the solution space for isothermal electrons. At $\mu=\mu_{c}$, there does not exist any solitary wave solution, the reasons of which being described in our theoretical section. The amplitude of PPSW decreases with increasing $\mu$ in $\mu_{p}\leq \mu < \mu_{c}$ and ultimately the PPSWs demolished at $\mu = \mu_{c}$. On the other hand, NPSWs come into the picture after $\mu$ crosses $\mu_{c}$ and the amplitude of NPSW gradually increases with $\mu$ in $\mu_{c}< \mu \leq \mu_{T}$. Thus one may assume that the point $\mu = \mu_{c}$ acts as a source for NPSWs and as well as a sink for PPSWs. There is no double layer solution for Maxwellian electrons. It is also important to note that if a NPSW exists for a given $\mu$ and for a given value of $\alpha$ at $\beta_{1}=0$, then the amplitude of this NPSW is the greatest among the amplitudes of all NPSWs at $M=M_{c}$ for the entire admissible range of $\beta_{1}$, i.e., the maximum amplitude NPSW at $M=M_{c}$ occurs when the nonthermal parameter $\beta_{1}$ assumes its smallest possible value 0. Again for $\mu_{r} \leq \mu \leq \mu_{T}$, the amplitude of NPSWs are not small.
  \item The amplitude of NPSW at $M=M_{c}$ decreases with increasing $\beta_{1}$ while the amplitude of PPSW at $M=M_{c}$ increases with increasing $\beta_{1}$.
  \item The amplitude of NPSW at $M=M_{c}$ increases with increasing $\mu$ while the amplitude of PPSW at $M=M_{c}$ decreases with increasing $\mu$.
  \item The entire numerical investigation depends on the solution space or compositional parametric space given by figure \ref{fig:sol space}. So for compactness of our investigation, it is customary to show that the solution space given by figure \ref{fig:sol space} is unique. To show that this solution space is unique we have to show that the values $\mu_{p}$, $\mu_{c}$, and $\mu_{r}$ maintain the same orderings for any value of $\alpha$ as they are in figure \ref{fig:sol space}. To do this, we have numerically investigated these values of $\mu$ and obtain figure \ref{fig:muprc}. This figure shows that for any value of $\alpha$, we always have $\mu_{p} < \mu_{c} <\mu_{r}$. Therefore, for any value of $\alpha$, the nature of solutions follow figure \ref{fig:sol space}.
\end{enumerate}


\bibliography{aps_adabkpd}

\begin{figure}
\begin{center}
  \includegraphics{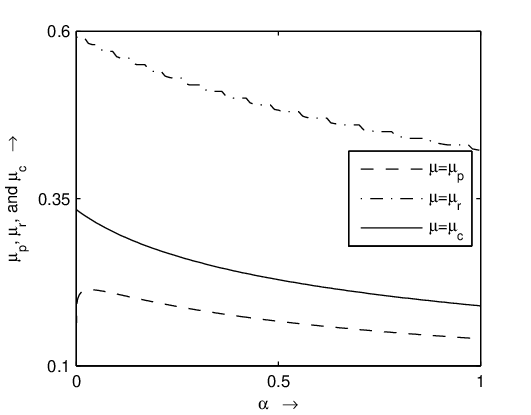}
  \caption{\label{fig:muprc} $\mu_{p}$, $\mu_{r}$ and $\mu_{c}$ are plotted against $\alpha$. Note that $\mu_{p} < \mu_{c} < \mu_{r}$ for any value of $\alpha$.}
\end{center}
\end{figure}

\begin{figure}
\begin{center}
  \includegraphics{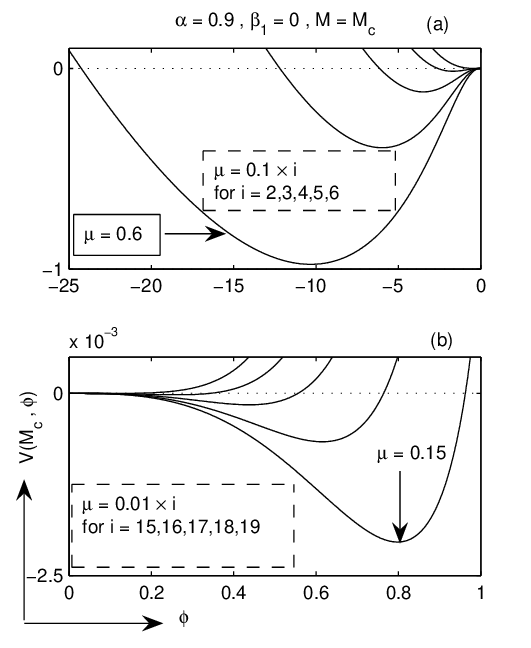}
  \caption{\label{fig:verify iso} $V(M_{c}, \phi)$ is plotted against $\phi$ for isothermal distribution of electrons ($\beta_{1}=0$) with values of $\mu$ lies within the intervals $\mu_{c} < \mu \leq \mu_{T}=0.6$ (figure (a)) and $\mu_{p} \leq \mu < \mu_{c}$ (figure (b)). For $\alpha = 0.9$, $\beta_{1}=0$, we have found $\mu_{p} \approx 0.145$ and $\mu_{c} \approx 0.195$.}
\end{center}
\end{figure}

\end{document}